\documentclass[aps,twocolumn,prb,superscript,floatfix,superscriptaddress,showpacs,footinbib,reprint]{revtex4-2}
\usepackage{hyperref}

\usepackage{dcolumn}                             %Align on the decimal point of numbers in tabular columns
\usepackage{bm}                                      %Access bold symbols in maths mode
\usepackage{graphicx}                             %Enhanced support for graphics
\usepackage{amsmath}                             %AMS mathematical facilities for LATEX - Used in this document for the 'aligned' environment
\usepackage{braket}			   %Dirac bra-ket and set notations
\usepackage{array}			   %Extending the array and tabular environments - Used in this document to increase/decrease dimensions on the 'table' environment
\usepackage{float}                                    %Improved interface for floating objects - Used in this document to place images and tables in the corresponding position of the text with the command [H] after the \begin{xxxx}
\usepackage{physics}                                %Macros supporting the Mathematics of Physics - Used in this document to correct Im{x} in math mode
\usepackage[titletoc]{appendix}               %Extra control of appendices
\usepackage{soul} % strikethrough, usage: \st{Hello World}

\usepackage[brazilian,british]{babel}      %Multilingual support for Plain TEX or LATEX
\usepackage{subfigure}  		            %Figures divided into subfigures
\usepackage[T1]{fontenc}                       %Standard package for selecting font encodings
\usepackage{ae}                                      %Virtual fonts for T1 encoded CMR-fonts
\usepackage[utf8]{inputenc}                   %Accept different input encodings

\usepackage{tikz}                                     %Create PostScript and PDF graphics in TEX
\usetikzlibrary{external}                           % set up externalization
\tikzset{external/system call={pdflatex \tikzexternalcheckshellescape -halt-on-error
-interaction=batchmode -jobname "\image" "\texsource" && 
pdftops -eps "\image.pdf"}}

\usepackage{gensymb}                           %Generic symbols for both text and math mode

\usepackage{booktabs}                            %Used for thicker rules in tables (\toprule, \midrule, etc)

\begin{document}

%\begin{frontmatter}
\title{The cumulant Green's functions method for the single impurity Anderson model}

%\author[Jorge]{J. H. Correa}
%\author[Figueira]{M. S. Figueira}
%\address[Jorge,Figueira]{Instituto de F\'{i}sica, Universidade Federal Fluminense, 24210-340, Niter\'oi, RJ, Brasil}

\author{T.~M.~Sobreira}
\affiliation{Instituto de F\'{i}sica, Universidade Federal Fluminense, Av. Litor\^anea s/N, 24210-340, Niter\'oi, RJ, Brazil}
\author{T.~O.~Puel}
\affiliation{Department of Physics and Astronomy, University of Iowa, Iowa City, Iowa 52242, USA}
\author{M.~A.~Manya}
\affiliation{Instituto de F\'{i}sica, Universidade Federal Fluminense, Av. Litor\^anea s/N, 24210-340, Niter\'oi, RJ, Brazil}
\author{S.~E.~Ulloa}
\affiliation{Department of Physics and Astronomy and Nanoscale and Quantum Phenomena Institute, Ohio University, Athens, Ohio 45701-2979, USA}
\author{G.~B.~Martins}
\affiliation{Instituto de F\'{i}sica, Universidade Federal de Uberl\^andia, Uberl\^andia, Minas Gerais, 38400-902, Brazil}
\author{J. Silva-Valencia}
\affiliation{Departamento de F\'{\i}sica, Universidad Nacional de Colombia, A. A. 5997 Bogot\'a, Colombia.}
\author{R.~N.~Lira}
\affiliation{Instituto de F\'{i}sica, Universidade Federal Fluminense, Av. Litor\^anea s/N, 24210-340, Niter\'oi, RJ, Brazil}
\author{M.~S.~Figueira}
\email{figueira7255@gmail.com}
\affiliation{Instituto de F\'{i}sica, Universidade Federal Fluminense, Av. Litor\^anea s/N, 24210-340, Niter\'oi, RJ, Brazil}

\begin{abstract}

Using the cumulant Green's functions method (CGFM), we study the single impurity Anderson model (SIAM). The CGFM starting point is a diagonalization of the SIAM Hamiltonian expressed in a semi-chain form, containing $N$ sites, viz., a correlated site (simulating an impurity) connected to the remaining $N-1$ uncorrelated conduction-electron sites. An exact solution can be obtained since the complete system has few sites. That solution is employed to calculate the atomic Green's functions and the approximate cumulants used to obtain the impurity and conduction Green's functions for the SIAM, and no self-consistency loop is required.

We calculated the density of states, the Friedel sum rule, and the impurity occupation number, all bench-marked against results from the numerical renormalization group (NRG). One of the main insights obtained is that,  at very low temperatures, only four atomic transitions contribute to generating the entire SIAM density of states, regardless of the number of sites in the chain and the model's parameters and different regimes: Empty orbital, mixed-valence, and Kondo. We also pointed out the possibilities of the CGFM as a valid alternative to describe strongly correlated electron systems like the Hubbard and $t-J$ models, the periodic Anderson model, the Kondo and Coqblin-Schrieffer models, and their variants.

\end{abstract}


\maketitle
%\end{frontmatter}

\section{Introduction}
\label{sec1}

In 1961, Anderson proposed a Hamiltonian~\cite{Anderson1961} to describe magnetic moment formation in metals containing diluted magnetic impurities~\cite{Friedel1958, Clogston1962}. That proposal became known as the single impurity Anderson model (SIAM). Band theory has proven its power in describing non-magnetic impurities diluted in semiconductors ~\cite{KohnLuttinger1955}, where the impurity introduces energy levels inside the band gap. However, more was needed to describe magnetic impurities diluted in a metallic matrix since the impurity orbitals, responsible for the net magnetic moment, have energies inside the host metal's conduction band. Anderson's central assumption was that the origin of the localized magnetic moments was the electronic  Coulomb correlation between electrons visiting the magnetic impurity inner orbitals. He solved the Hamiltonian through a Hartree-Fock approximation and obtained the model's non-magnetic/magnetic phase diagram.

The subsequent critical discovery in the study of magnetic impurities diluted in metallic hosts came from the experiments of Myriam Sarachik \emph{et al.}~\cite{Sarachik1964}, who measured the resistivity as a function of temperature in a series of $\rm{Mo-Nb}$ and $\rm{Mo-Re}$ alloys, with and without $1\%$ Fe, at room temperature, and between $1.5$ and $77$ K. Those measurements revealed the existence of a resistivity minimum, which was associated with the localized magnetic moments of the impurities. In such systems, the resistivity decreases monotonically with temperature up to a specific characteristic temperature, below which it increases again with an $ln (T)$ behavior. 

After entering in contact with Sarachik's experimental results, J. Kondo showed, employing second-order perturbation theory, that the resistivity minimum was produced by spin-flip scattering of conduction electrons by the localized magnetic moments~\cite{Kondo1964}. This scattering process competes with the lattice vibrations of the host lattice and only becomes dominant at very low temperatures. Kondo obtained that the contribution of the spin-flip scattering to the resistivity is given by $J^{3}ln (T)$, where $J$ is the exchange coupling between a localized magnetic moment and the conduction electrons. For $J < 0$, the system is dominated by antiferromagnetic correlations and below the temperature that defines the resistivity minimum, known as the Kondo temperature $T_K$; this term diverges, indicating a non-physical logarithmic divergence. That only indicates that the perturbation theory is no longer valid below $T_K$. The Hartree-Fock solution obtained by Anderson~\cite{Anderson1961} could not describe the resistivity minimum. Systems that exhibit spin-flip scattering are known as Kondo systems. The genuine contribution of Myrian Sarachick to the discovery of the Kondo effect can be found in a paper on her memories ~\cite{Sarachik2018}.

The next important step, on the theoretical side,  was given by Yamada~\cite{Yamada75}, which, using a perturbative expansion (up to fourth order in powers of the local electronic correlation $U$), obtained that the $T=0$ impurity density of states consists of three peaks, viz., two broad structures located at $E_{d}$ and $E_{d}+U$ ($E_d$ is the impurity orbital energy), and a narrow peak located at the Fermi level, related directly to the Kondo effect. In the same year as the appearance of Yamada's work, Kenneth Wilson published his seminal work on Renormalization techniques applied to the SIAM \cite{Wilson1975}: ``The renormalization group: Critical phenomena and the Kondo problem", which is a numerical technique that describes accurately the rising of the Kondo peak in the Kondo limit of the model. Following this work, intense research improves the application of the method; some of those works are the references, Refs.~\cite{Krishna-murthy1980, Bulla2008, Costi2014}. The increased computer power over the years turns the method into a standard for Kondo impurity systems.

In two recent papers~\cite{Lira_JPCM_2023, Lira_PLA_2023}, three of the authors (R. N. Lira, J. Silva-Valencia, and M. S. Figueira) developed the Cumulant Green's Function Method (CGFM) for the single band Hubbard model. In this paper, we extend its derivation to the SIAM. In Appendix \ref{sec6},  we briefly review the CGFM fundamental relations and detail the basic ideas for calculating the approximate full SIAM GFs of the CGFM in four steps: (i) choice of a cluster, impurity plus conduction band sites, to be solved employing exact diagonalization (ED) methods; (ii) using the Lehmann representation, to calculate all the atomic Green's functions associated to the possible transitions inside the cluster's Hilbert space; (iii) employing those atomic GFs, the atomic cumulants are calculated; (iv) The atomic cumulants obtained in the previous step are used to calculate the impurity and conduction GFs for the SIAM. The CGFM can also be applied to other strongly correlated electron systems like the periodic Anderson model (PAM), the $t-J$, Kondo, and Coqblin-Schrieffer models and their variants. We employ the NRG Ljubljana open source code \cite{Zitko2021}  to benchmark the results obtained with the CGFM.

To obtain the atomic cumulants, we calculate all the atomic transitions inside the Hilbert space generated by the cluster solution defined in step (i) until the available computational limit is reached. At first sight, it seems that the CGFM is only a brute force method; however, using this procedure, we discovered that, unexpectedly, at low temperatures, independent of the parameters and regime (empty orbital, mixed-valence, or Kondo \cite{Costi2010}), only four transitions, out of the vast Hilbert space produced in the ED calculation, contribute to generating the whole SIAM density of states (two transitions associated with the Hubbard sub-bands and two with the Kondo peak), including the Kondo region, which we study in detail. We consider this result to be the main achievement of the work, and we will discuss this striking point and its consequences in the following sections.

This work has the following structure: In Sec.~\ref{sec2}, we present the SIAM Hamiltonian and discuss its transformation to a finite semi-chain containing $N$ sites and the strategies employed by the CGFM to solve this problem. In Sec.~\ref{sec3}, we introduce, in four steps, the basic ideas of the method. In Sec. \ref{sec4}, the results obtained with the CGFM,  the density of states with the corresponding Kondo peak formation, the Friedel sum rule, and the occupation numbers are benchmarked against the corresponding ones obtained by NRG. In Sec.~\ref{sec5}, we present our conclusions and outline further possible developments and applications of the CGFM. Finally, in Appendix \ref{sec6},  we briefly review the CGFM fundamental relations.

\section{The single impurity Anderson model: The CGFM strategy}
\label{sec2}

\begin{figure}[h]
\centering
\includegraphics[clip,width=0.45\textwidth,angle=0.]{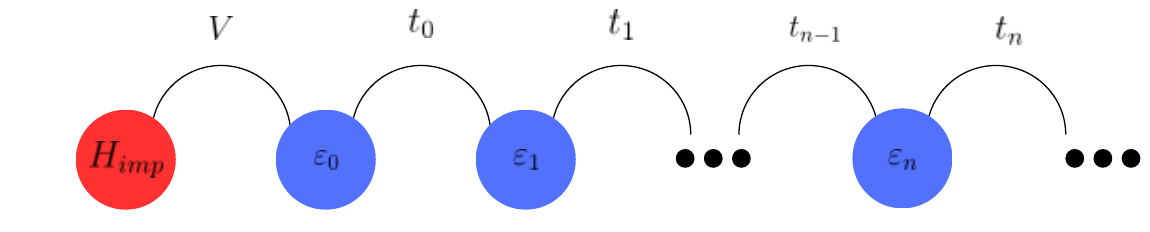}
	\caption{The finite atomic linear chain (FALC) form of the SIAM (Eq. \ref{Linearchain}), which the CGFM uses.}
	\label{LChain}
\end{figure}

Here, we briefly discuss how NRG and CGFM solve the SIAM, highlighting where the methods are similar and diverge from each other. Both methods use a finite chain to model the conduction band. The NRG solves the SIAM via Wilson's method that maps a logarithmically discretized conduction band into a one-dimensional chain ~\cite{Krishna-murthy1980, Bulla2008, Costi2014}. Because of the logarithmic discretization, the hopping parameters between neighboring sites fall off exponentially. The CGFM resorts to a real-space tight-binding finite chain, as depicted in Fig.~\ref{LChain}, with all hoppings being a constant along the chain, i.e.,  $t_n=1$ for all $n=1 \cdots N$, with $N$ being the number of sites considered in the chain. We call it the finite atomic linear chain (FALC) and then use N-FALC to specify the number of sites employed in the calculations. The SIAM~\cite{Anderson1961} Hamiltonian is given by
\begin{align}
&H_{\rm SIAM} =\sum_{\sigma}\epsilon_{f} n_{f\sigma}+Un_{f\uparrow}n_{f\downarrow}+\nonumber \\
&\sum_{k\sigma} V_{k} ({c}^{\dagger}_{k\sigma}{f}_{\sigma}+{f}^{\dagger}_{\sigma}{c}_{k\sigma})+
\sum_{k\sigma} \epsilon_{k}{c}^{\dagger}_{k\sigma}{c}_{k\sigma}, 
\label{AIM}
\end{align}
\noindent
where ${f^{\dagger}_{\sigma}}$ (${f_{\sigma}}$) creates (annihilates) an electron at the impurity, while $n_{f\sigma}={f}^{\dagger}_{\sigma}{f_{\sigma}}$ is the impurity's electron-number operator. The operator ${c^{\dagger}_{k\sigma}}$ (${c_{k\sigma}}$) 
creates (annihilates) conduction electrons, where $\sigma$ labels the spin, which may point up or down ($\uparrow$, $\downarrow$), respectively. The first two terms represent the impurity part of the Hamiltonian, with $\epsilon_{f}$ being the impurity orbital energy $E_{f}$, measured from the chemical potential $\mu$ ($\epsilon_{f}=E_{f}-\mu$) and the second term, 
represents the Coulomb repulsion $U>0$ between two electrons occupying the impurity. The third term describes the coupling between the impurity and the conduction electrons through the hybridization $V_{k}$. Finally, the fourth term describes the conduction electrons kinetic energy, which is characterized by single-particle energies $E_{k}$, with a dispersion relation $\epsilon_{k}=E_{k}-\mu$. 
\begin{center}
\begin{table}[tbh]
\begin{tabular}{|l|l|l|l|l|}
\hline
$I_{x}$ & $1$ & $2$ & $3$ & $4$ \\ \hline
$\alpha=(b,a)$ & $(0,+)$ & $(0,-)$ & $(-,d)$ & $(+,d)$ \\ \hline
\end{tabular}%
\caption[TABLE I]{ Index representation of the possible transitions in the finite $U$ SIAM. }
\label{table1}
\end{table}
\end{center}

The impurity, in the finite $U$ SIAM, has four possible states, viz., empty ($\ket{0}$), spin-up ($\ket{+}$), spin-down ($\ket{-}$), and doubly occupied ($\ket{d}$), with four possible transitions between those states, as described in Table~\ref{table1}, where we employ the index $\alpha=(b, a)$ to label each of those transitions, with the meaning $(b, a) \equiv \ket{a} \rightarrow \ket{b}$. For example, transitions $I_{x}=1$, and $3$ annihilate a spin-up electron at the impurity, while transitions $2$ and $4$ annihilate a spin-down electron. For simplicity, we did not represent in Table~\ref{table1} the reversed transitions that create an electron at the impurity, but they are included in all the calculations.

In the development of the CGFM, we had to address two problems. The first one was how to derive the particular relation between the GFs, $G_{\sigma }(k,z)$ and the cumulants $M_{\sigma }^\text{eff}(k,z)$. The second was to find out how to calculate the cumulants themselves. The first had already been solved in earlier papers~\cite{Figueira1994, Foglio2010} that calculated the formal exact relation between the cumulants and Green's functions (GFs), valid for all dimensions. In the GF derivation, we used Hubbard-operator diagrams to rearrange all the cumulants that contribute to the exact impurity GF, $G^{f}_{\sigma }(k,z)$. We defined an effective cumulant $M_{\sigma }^\text{eff} (k,z)$, which can be obtained through all the diagrams in $G^{f}_{\sigma }(k,z)$ that cannot be separated by cutting a single edge (usually called ``proper'' or ``irreducible'' diagrams). The exact $G^{f}_{\sigma }(k,z)$ is then given by a family of chain diagrams, but with the effective cumulant $M_{\sigma }^\text{eff}(k,z)$ dressed by correlations, given by
\begin{equation}
G^{f}_{\sigma }(k,z)=\frac{M_{\sigma }^{eff}(k,z)}{1-\mid V({\bf k})\mid ^{2}G^{0c}_{\sigma }(k,z) M_{\sigma }^{eff}(k,z)},  \label{E2.8a}
\end{equation}
where $z=i\omega_n=(2n+1)\pi k_{B}T$ ($n=\pm 1, \pm 2,...$) correspond to Matsubara frequencies along the imaginary axis.
$G^{0c}_{\sigma }(k,z) =-1/(z-\varepsilon(k))$, where $\varepsilon(k)$ is the uncorrelated dispersion relation. Similarly, one obtains the exact GF for the conduction electrons, namely
\begin{equation}
G^{c}_{\sigma }(k,z) =\frac{-\ 1}{z-\varepsilon (k)+\mid V(k)\mid ^{2}M_{\sigma }^{eff}(k,z)} .  \label{E2.8b}
\end{equation}

The second problem requires an additional explanation. The conventional way of calculating cumulants is via perturbative expansions employing Feynmann or Hubbard-operator diagrams~\cite{Metzner1991, Figueira1994}. However, this procedure suffers from several limitations when one tries to go beyond well-established approximations such as {\it Hartree-Fock}, {\it ladder}, or {\it ring} approximations~\cite{Fetter1971}. Indeed, when pushed beyond those standard approximations, the mathematical complexity
increases dramatically, and it is usual to obtain GFs displaying pathological behaviors, such as parameter regions where the density of states or the entropy are negative~\cite{Roura-Bas2016}.

To overcome the limitations mentioned above, in a previous  work~\cite{Lobo2010}, we developed an alternative way to obtain the
cumulants, employing the solution for a subset of sites of the original Hamiltonian. We replace $M_{\sigma }^\text{eff}(k,z)$ by the corresponding quantity for an exactly solvable model,  which is the atomic limit of the SIAM that consists of a dimer, i.e., the impurity and one conduction site. In the present paper, we generalize that procedure by modeling the SIAM in the form of a finite atomic linear chain (FALC) containing $N$ sites, as indicated in Eqs.~\eqref{Tridiag} and \eqref{Linearchain} \cite{Costi2014}, and represented pictorially in Fig.~\ref{LChain}. The chain's first site corresponds to the impurity; the remaining are $N-1$ conduction electron sites. Because of the lack of translation symmetry, the approximate effective cumulant $M_{\sigma }^\text{eff}(k,z)=M_{\sigma }(z)$ obtained is independent of wave vector ${\bf k}$. The more sites we include in the N-FALC, the more atomic transitions are included in the calculations, and the better we describe the approximate cumulant $M_{\sigma }(z)$. In what follows, we always consider the cumulants  $M_{\sigma }(z)$, independent of the wave vector $\vec{k}$. 

The conduction band can be discretized according to the relation~\cite{Costi2014}
\begin{equation}
\sum_{k\sigma} \epsilon_{k}{c}^{\dagger}_{k\sigma}{c}_{k\sigma} \rightarrow 
\sum_{n=0,\sigma}^{\infty}  [\epsilon_{n}{c}^{\dagger}_{n\sigma}{c}_{n\sigma}+
t_{n} ({c}^{\dagger}_{n\sigma}{c}_{n+1\sigma}+{c}^{\dagger}_{n+1\sigma}{c}_{n\sigma})].
\label{Tridiag}
\end{equation}
Substituting Eq.~\eqref{Tridiag} into Eq.~\eqref{AIM}, we obtain 
\begin{align}
H_\text{SIAM} =&\sum_{\sigma}\epsilon_{f} n_{f\sigma}+Un_{f\uparrow}n_{f\downarrow}+
V\sum_{\sigma} ({c}^{\dagger}_{0\sigma}{f}_{\sigma}+{f}^{\dagger}_{\sigma}{c}_{0\sigma})+\nonumber \\
&\sum_{n=0,\sigma}^{\infty}  [\epsilon_{n}{c}^{\dagger}_{n\sigma}{c}_{n\sigma}+
t_{n} ({c}^{\dagger}_{n\sigma}{c}_{n+1\sigma}+{c}^{\dagger}_{n+1\sigma}{c}_{n\sigma})], 
\label{Linearchain}
\end{align}
where, in the calculation of the N-FALC atomic cluster, we use a constant hybridization  $V$, the hopping $t_{n}=t$, and the conduction sites energies $\epsilon_{n}=0$.

\section{The cumulant Green's function method: Some useful relations}
\label{sec3}

After obtaining the SIAM GFs in Appendix \ref{sec6}, Eqs. \ref{E5.12}-\ref{E5.36}, we can calculate the dynamical properties of the system using the localized and conduction density of states (DOS) that are given by
\begin{equation}
\rho^{(f,c)}_{\sigma}(\omega )=\left( \frac{-1}{\pi }\right)
Im[G^{(f,c)}_{\sigma}(\omega )].  \label{E6.1}
\end{equation}
We can also calculate the occupation numbers $n^{\sigma}$ for the spin up and down. They should satisfy the completeness relation per spin
\begin{equation}
comp.=n^{\sigma}_{vac}+n^{\sigma}_{up}+n^{\sigma}_{down}+n^{\sigma}_{d}=1,  \label{Occ1}
\end{equation}
with $\sigma=\uparrow$ representing the transitions associated with indices 1 and 3, and $\sigma=\downarrow$ associated with indices 2 and 4 (see Table \ref{tab:3} in the Appendix \ref{sec6}). The first term of Eq. \ref{Occ1} is the vacuum occupation number, the second and the third terms are the occupation of the spin up and down, respectively, and the last is the double occupation. All the different averages can be calculated employing Green's functions $G_{1}(\omega )$, $G_{3}(\omega )$, $G_{2}(\omega )$ and $G_{4}(\omega )$ defined by equations (\ref{E5.12}), and (\ref{E5.13}):
\begin{equation}
n_ {\text{vac}}=\left( \frac{1}{\pi }\right)
\int_{-\infty }^{\infty }\text{d}\omega Im(G_{1})(1-n_{\text{F}})+Im(G_{2})(1-n_{\text{F}}),  \label{G001}
\end{equation}%
\begin{equation}
n_{\text{up}}=\left( \frac{1}{\pi }\right)
\int_{-\infty }^{\infty }\text{d}\omega Im(G_{1})n_{\text{F}}+Im(G_{4})(1-n_{\text{F}}),  \label{Gup1}
\end{equation}%
\begin{equation}
n_ {\text{down}}=\left( \frac{1}{\pi }\right)
\int_{-\infty }^{\infty }\text{d}\omega Im(G_{3})(1-n_{\text{F}})+Im(G_{2})n_{\text{F}},  \label{Gdown1}
\end{equation}%
\begin{equation}
n_ {\text{d}}=\left( \frac{1}{\pi }\right)
\int_{-\infty }^{\infty }\text{d}\omega Im(G_{3})n_{\text{F}}+Im(G_{4})n_{\text{F}},  \label{Gdd1}
\end{equation}%
where $n_{\text{F}}(x)=1/\left[ 1+\exp (\beta x)\right]$ is the Fermi-Dirac distribution. 

The electron density per lattice site $n$  (electron concentration or band filling) is defined according to equations 
(\ref{Gup1})-(\ref{Gdd1}) as
\begin{equation}
n=\frac{N_{\text{e}}}{N} =n_{\text{up}}+n_ {\text{down}}+2n_ {\text{d}},
\label{Density}
\end{equation}
where $N_{\text{e}}$ is the number of electrons and $N$ is the total number of lattice sites. The factor $2$ in front of $n_{\text{d}}$ refers to the number of electrons inside the double-occupied state. The maximum number of electrons per site in the SIAM is $2$, and the cases $n = 0$ and $n = 1$ correspond to empty and half-filled bands, respectively.

In this work, we will use as the reference for the Kondo temperature ${T_{K}}$, the result obtained by Haldane \cite{Haldane1978} in the wide-band limit
\begin{equation}
\label{eq:hald1}
	{T}_{K} =0.364 \left(\frac{2{\Delta} U}{\pi}\right)^{\frac{1}{2}}
	\exp\left[\frac{\epsilon_f\left(\epsilon_f+U \right)}{2{\Delta} U/\pi}\right](1+O(\Delta)).
\end{equation} 

On the other hand, we will employ the halfwidth taken at half maximum (HWHM), $\Gamma_{K}^{0}$  \cite{Jacob2023, Turco2023} to
define the Kondo peak width. In particular, we will use the results found numerically by Zitko and Pruschke from NRG calculations \cite{Zitko2009} for the symmetric case of the SIAM to relate the Haldane's $T_{K}$ with  $\Gamma_{K}^{0}$.
\begin{equation}
\Gamma_{K}^{0} \simeq 3.7 T_{K} .
\label{TKZitko}
\end{equation}

For a given set of parameters, we use Eq. \ref{eq:hald1} to calculate Haldane's $T_{K}$ and then use Eq. \ref{TKZitko} to find the Kondo peak HWHM, $\Gamma_{K}^{0}$.

Another equation that needs to be satisfied is the Friedel Sum Rule (FSR) (Friedel, 1958). In the context of the SIAM, the additional states created below the Fermi energy correspond to the occupation numbers of the localized states, and the scattering potential represents the hybridization effect on the conduction electrons. Thus, the FSR can be expressed as \cite{Langreth1966}
\begin{equation}
\rho _{f\sigma }(\mu )=\frac{sin^{2}\left( \pi n_{f\sigma }\right) }{\Delta
\pi },  \label{fried}
\end{equation}%
where $\rho _{f,\sigma }(\mu )$ is the density of states of the localized level at the chemical potential. To be rigorous, the FSR is only valid at $T=0$, but it still can be used as an approximation at very low temperatures below the $T_K$.

\section{Results and discussion}
\label{sec4}

In this Section, we present the results of the CGFM for the SIAM, and we benchmarked all the results against the NRG Ljubljana open source code \cite{Zitko2021}. We set all units to the Anderson parameter $\Delta=0.01D$, with $D=1$ being the semi-width of the uncorrelated conduction band, given by Eq. \ref{SB1}, and fixed the system at the chemical potential $\mu=0$, and temperature $T=0.0001\Delta$, below the Haldane's Kondo temperature, Eq. \ref{eq:hald1}.

\begin{figure}[h]
\centering
\includegraphics[clip,width=0.23\textwidth,angle=0.]{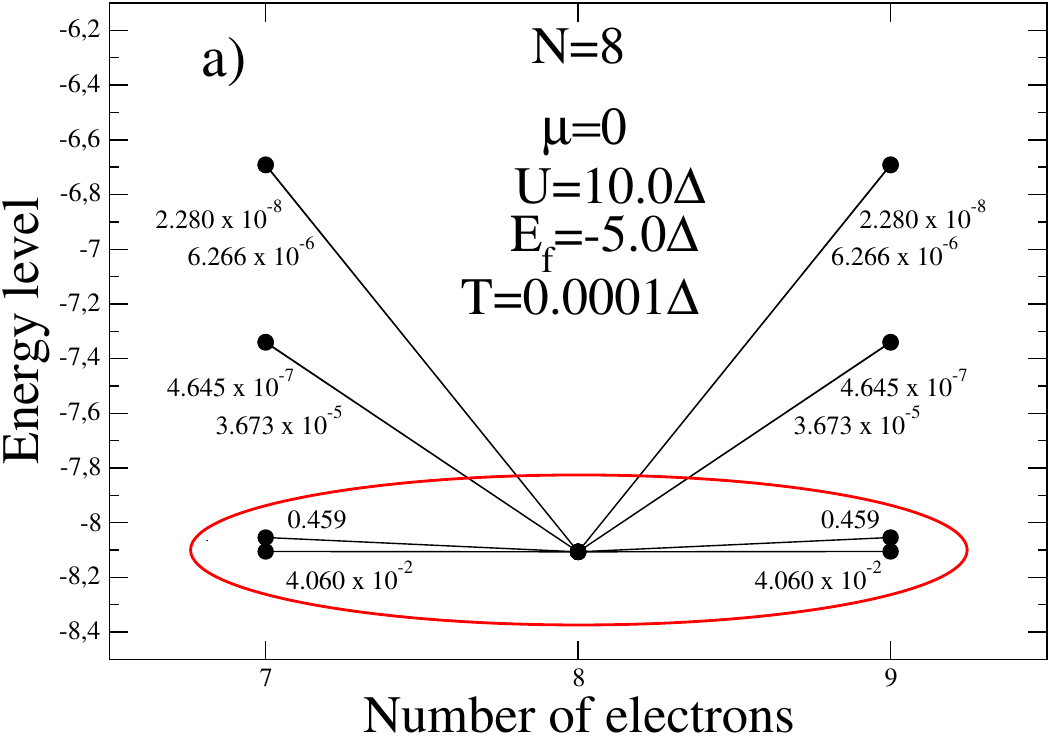}
\includegraphics[clip,width=0.23\textwidth,angle=0.]{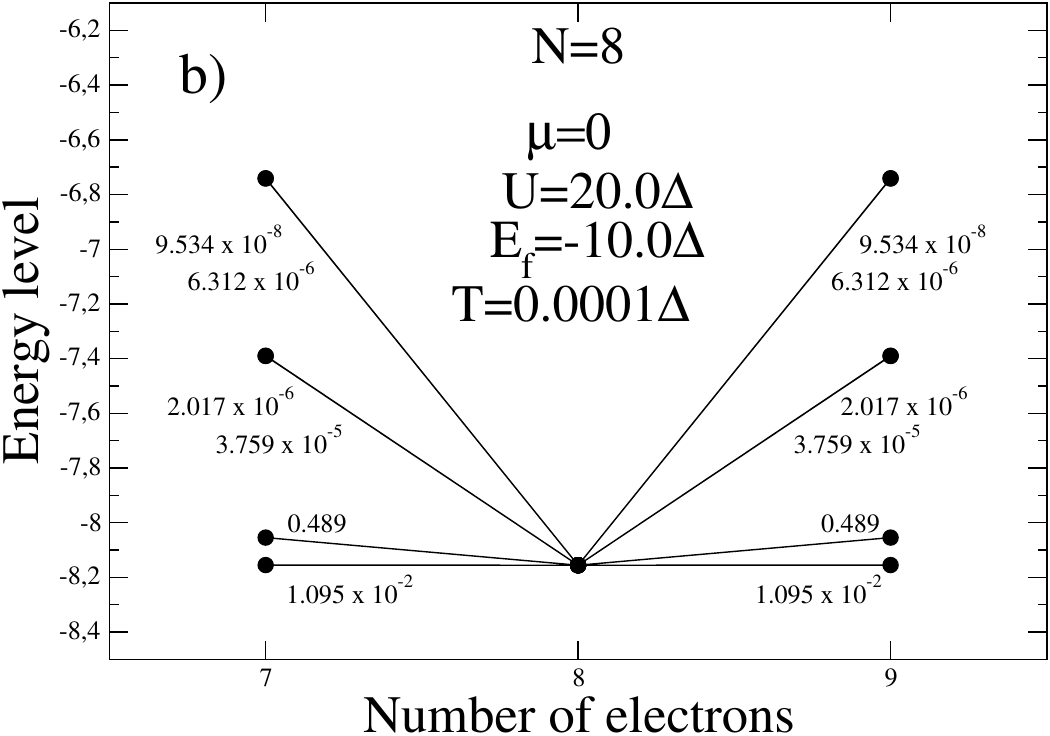}\\
\includegraphics[clip,width=0.23\textwidth,angle=0.]{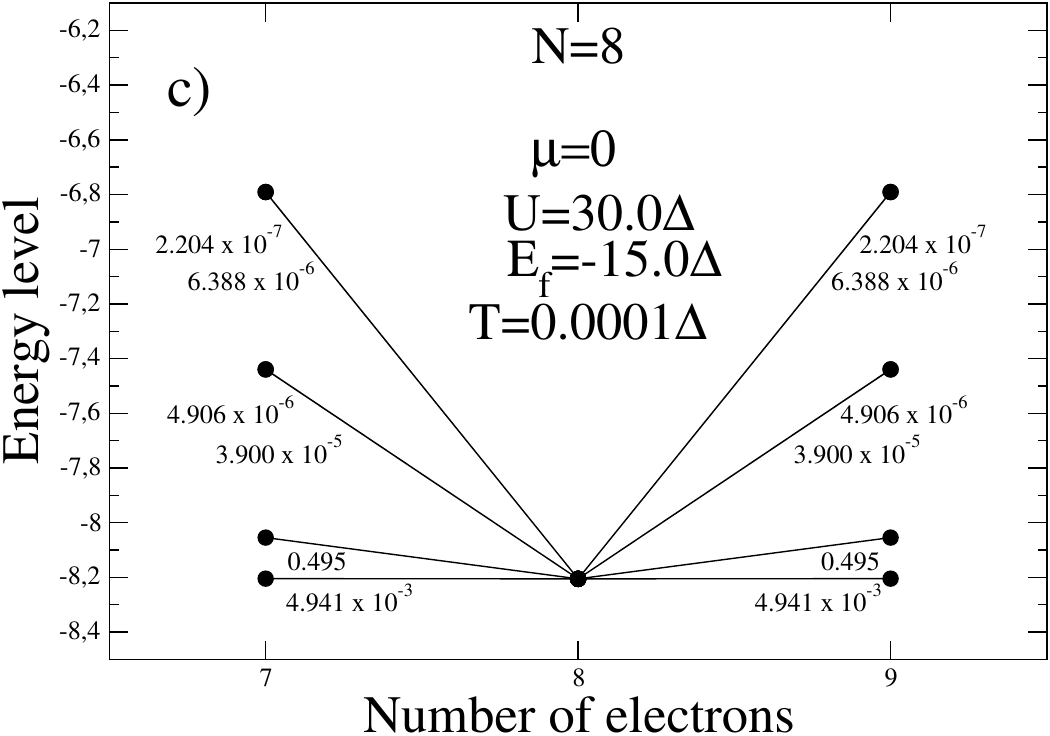}
\includegraphics[clip,width=0.23\textwidth,angle=0.]{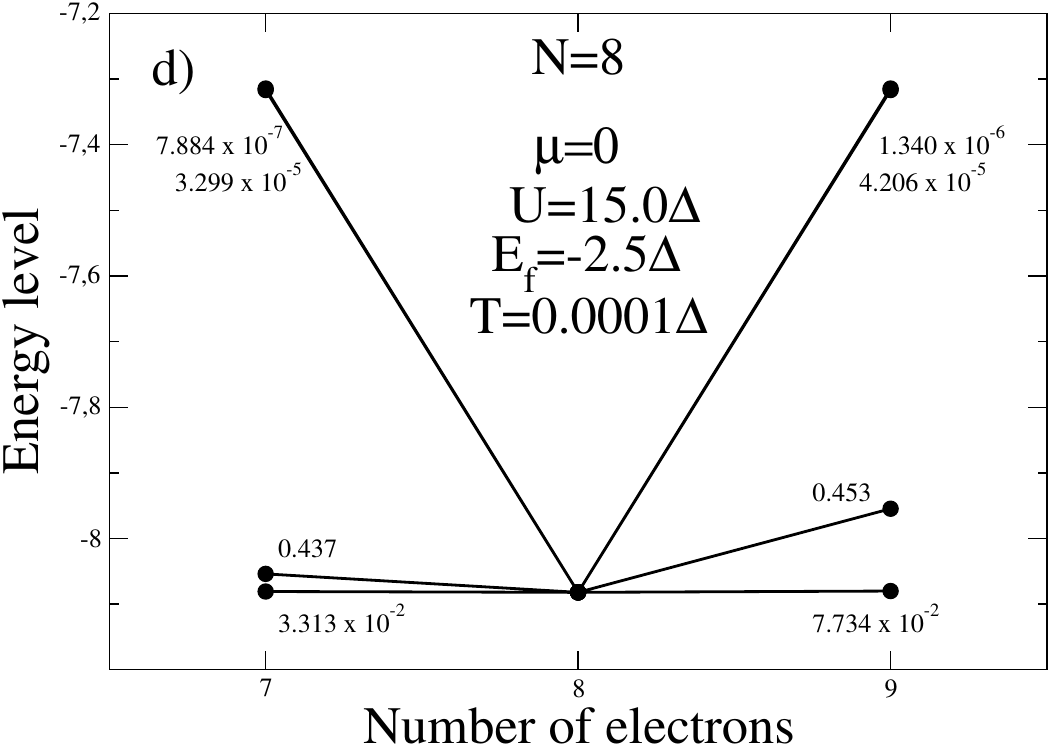}\\
\includegraphics[clip,width=0.23\textwidth,angle=0.]{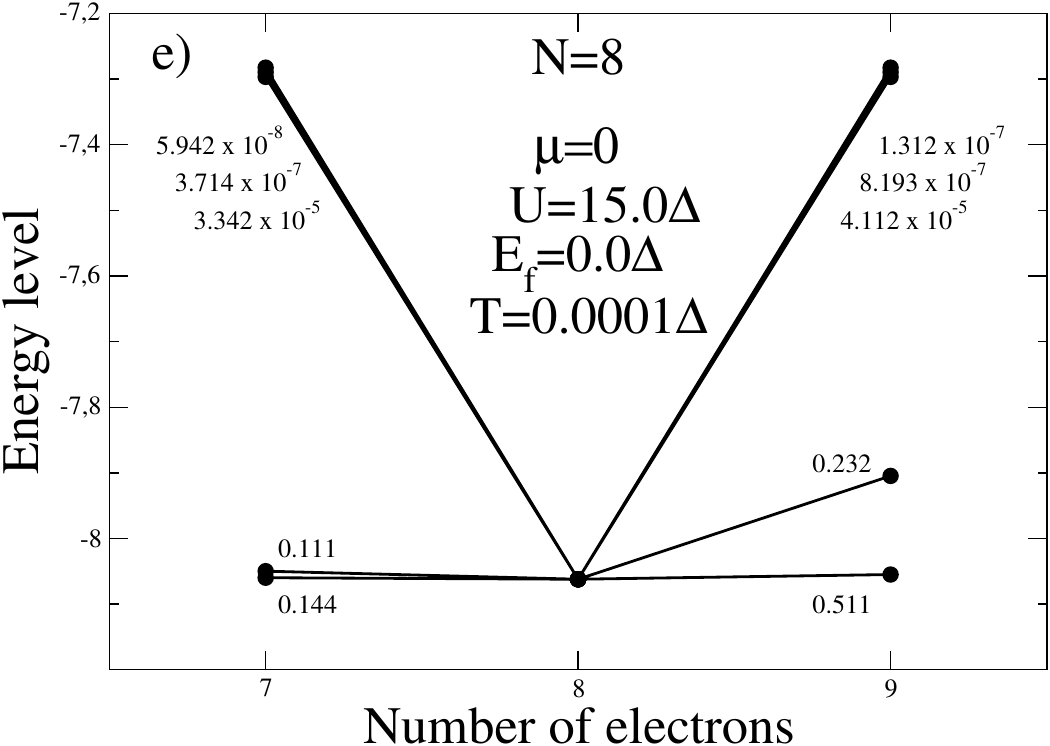}
\includegraphics[clip,width=0.23\textwidth,angle=0.]{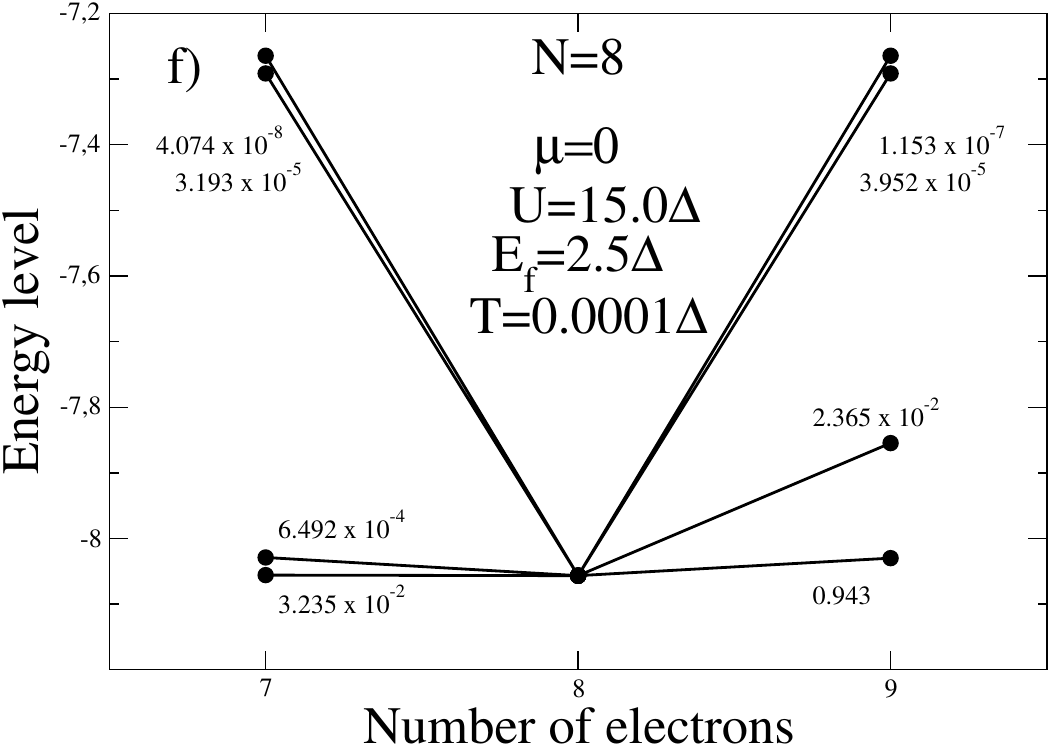}
	\caption{At low temperatures and for $N=8$, possible transitions are shown for the particle-hole symmetric case (a)-(c) and the asymmetric case (d)-(f), representing the regimes of mixed-valence/Kondo (d), mixed-valence (e), and empty orbital (f) respectively.}
\label{transitions}
\end{figure}
\begin{figure}[th]
\centering
\includegraphics[clip,width=0.45\textwidth,angle=0.]{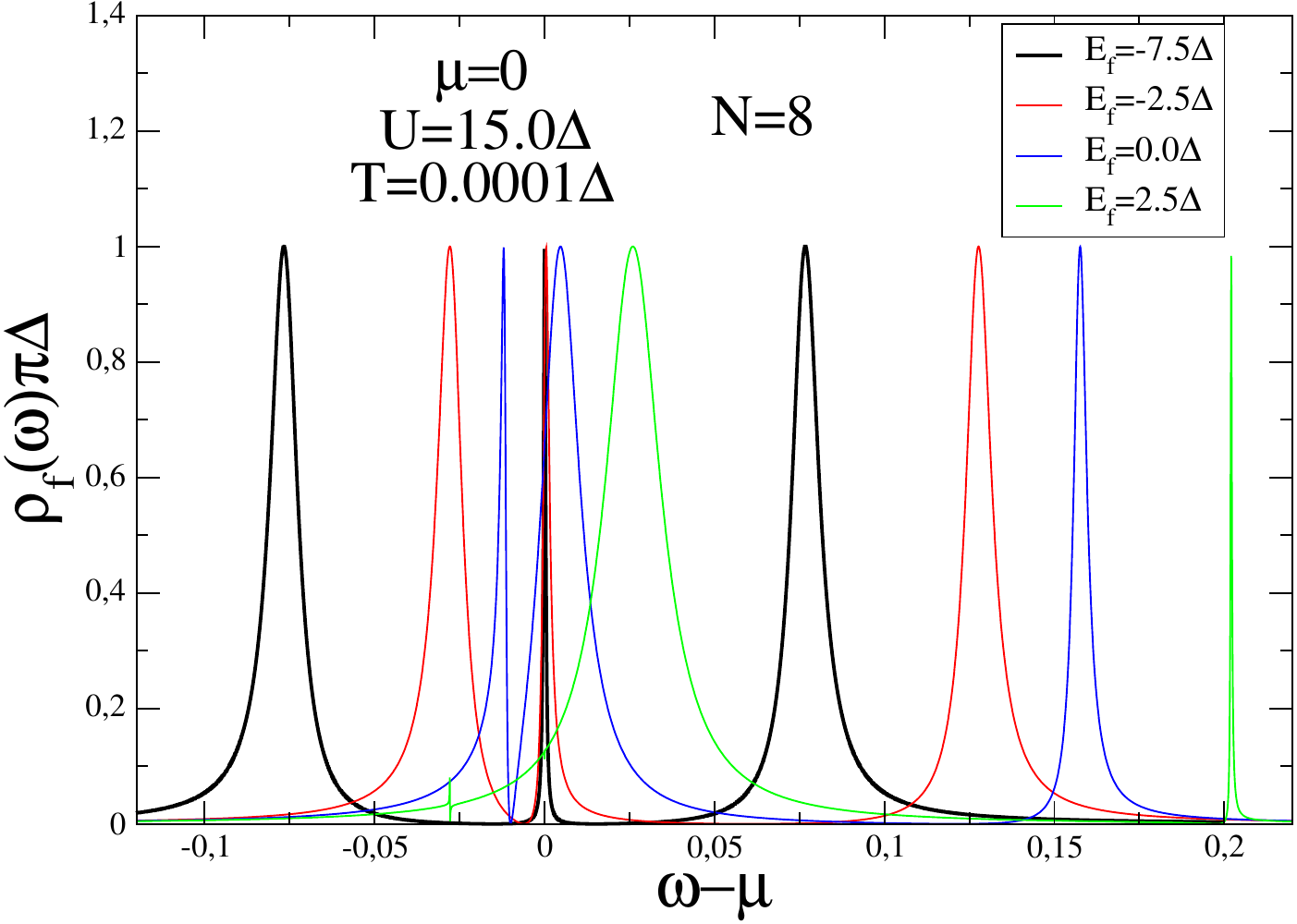}
\caption{The density of states for different regimes of the SIAM, at low temperature, with the corresponding occupation numbers given in Fig. \ref{fig4}. The colors refer to different regimes: empty orbital regime (green curve), mixed-valence regime (blue curve), mixed-valence/Kondo regime (red curve), and Kondo regime (black curve).}
\label{Kondo_formation}
\end{figure}
\begin{figure}[th]
\centering
\includegraphics[clip,width=0.23\textwidth,angle=0.]{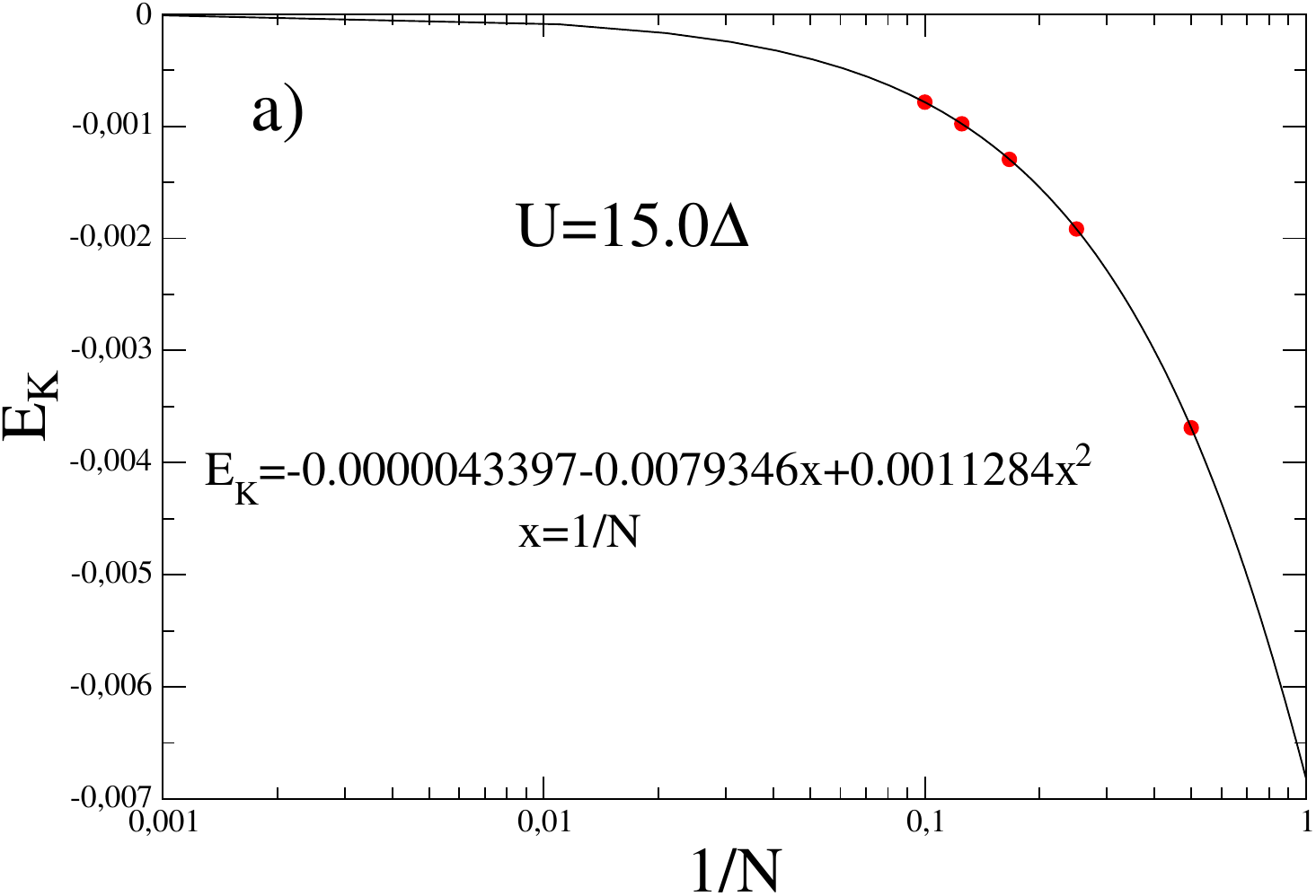}
\includegraphics[clip,width=0.23\textwidth,angle=0.]{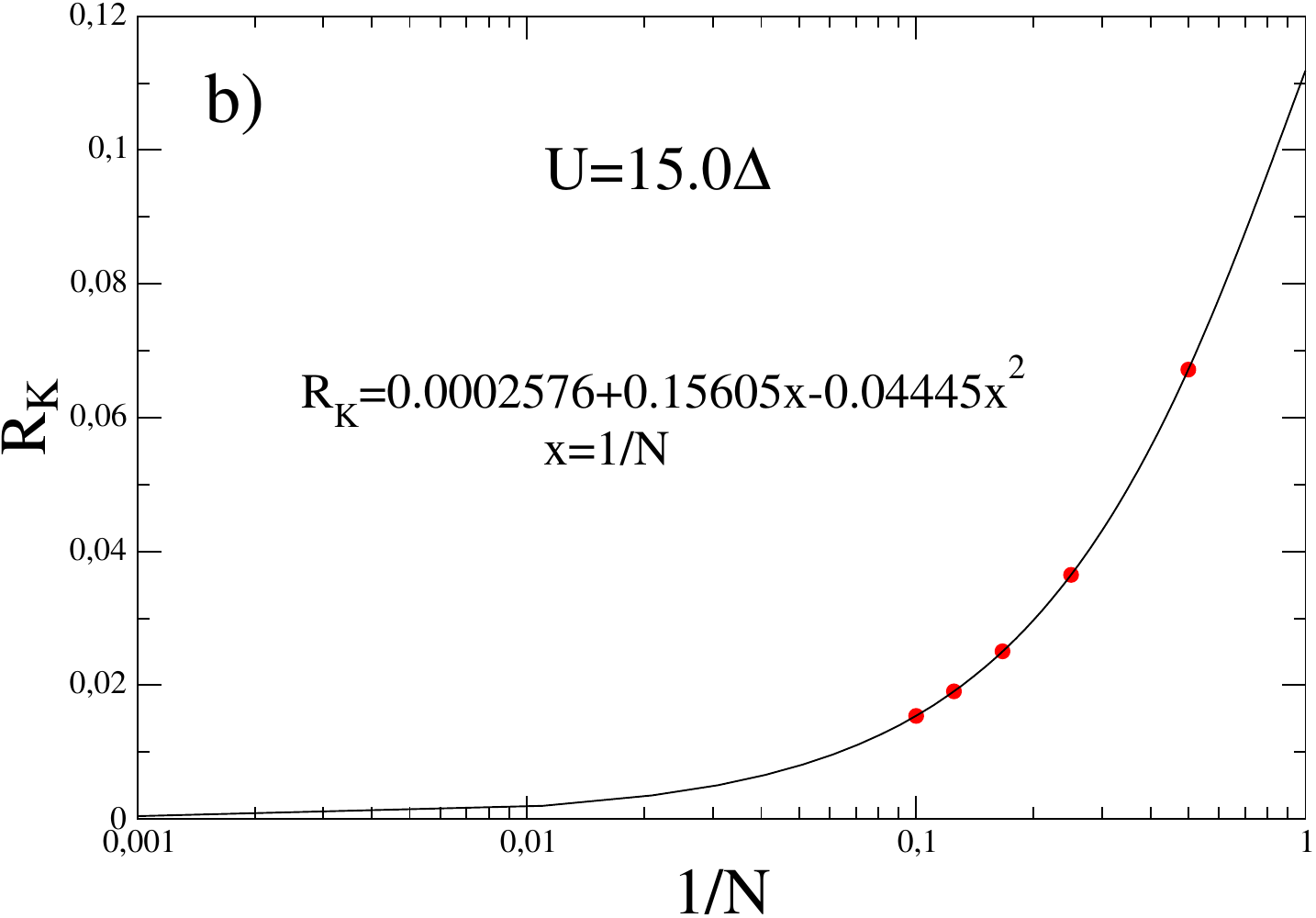} \\
\includegraphics[clip,width=0.23\textwidth,angle=0.]{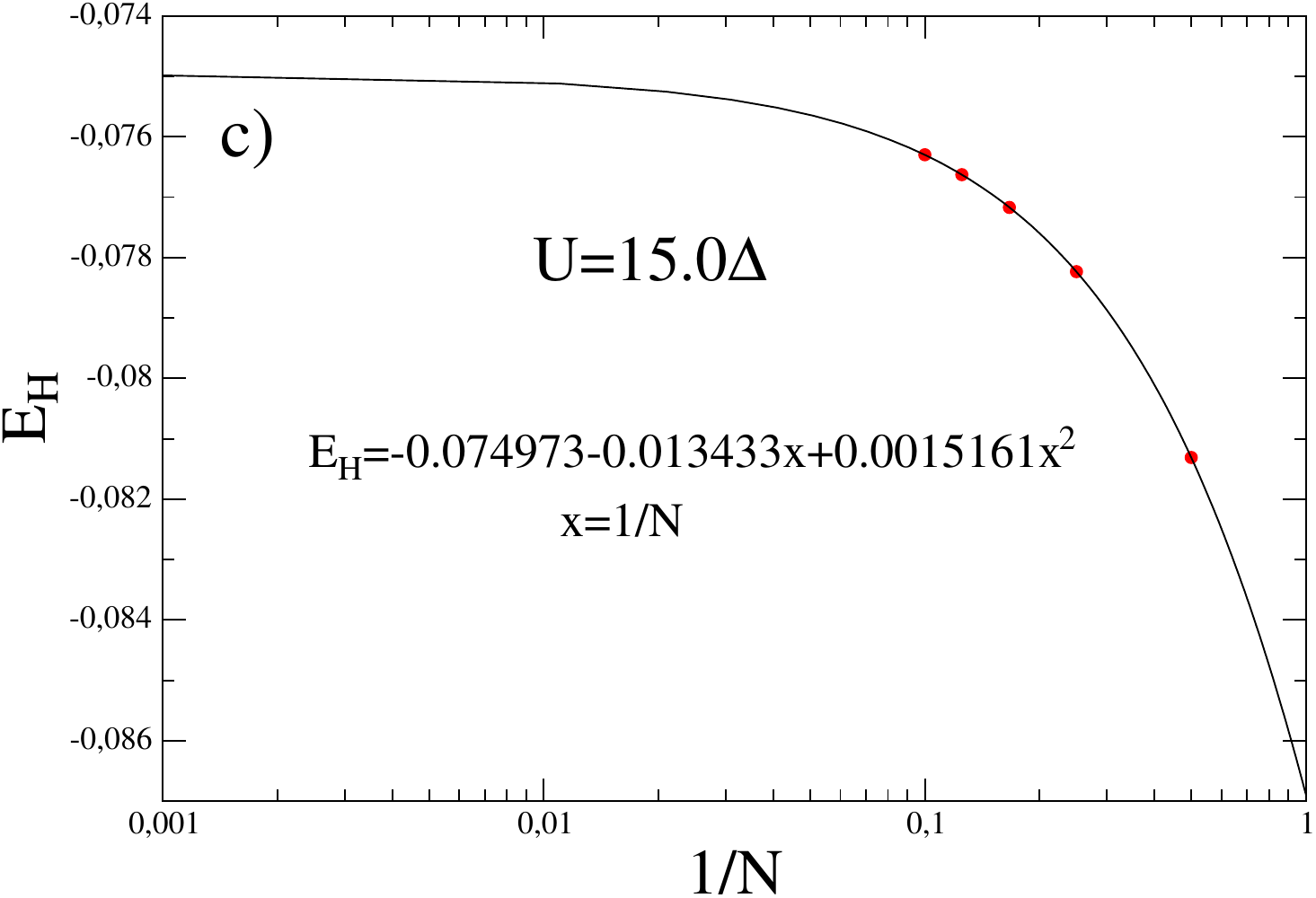}
\includegraphics[clip,width=0.23\textwidth,angle=0.]{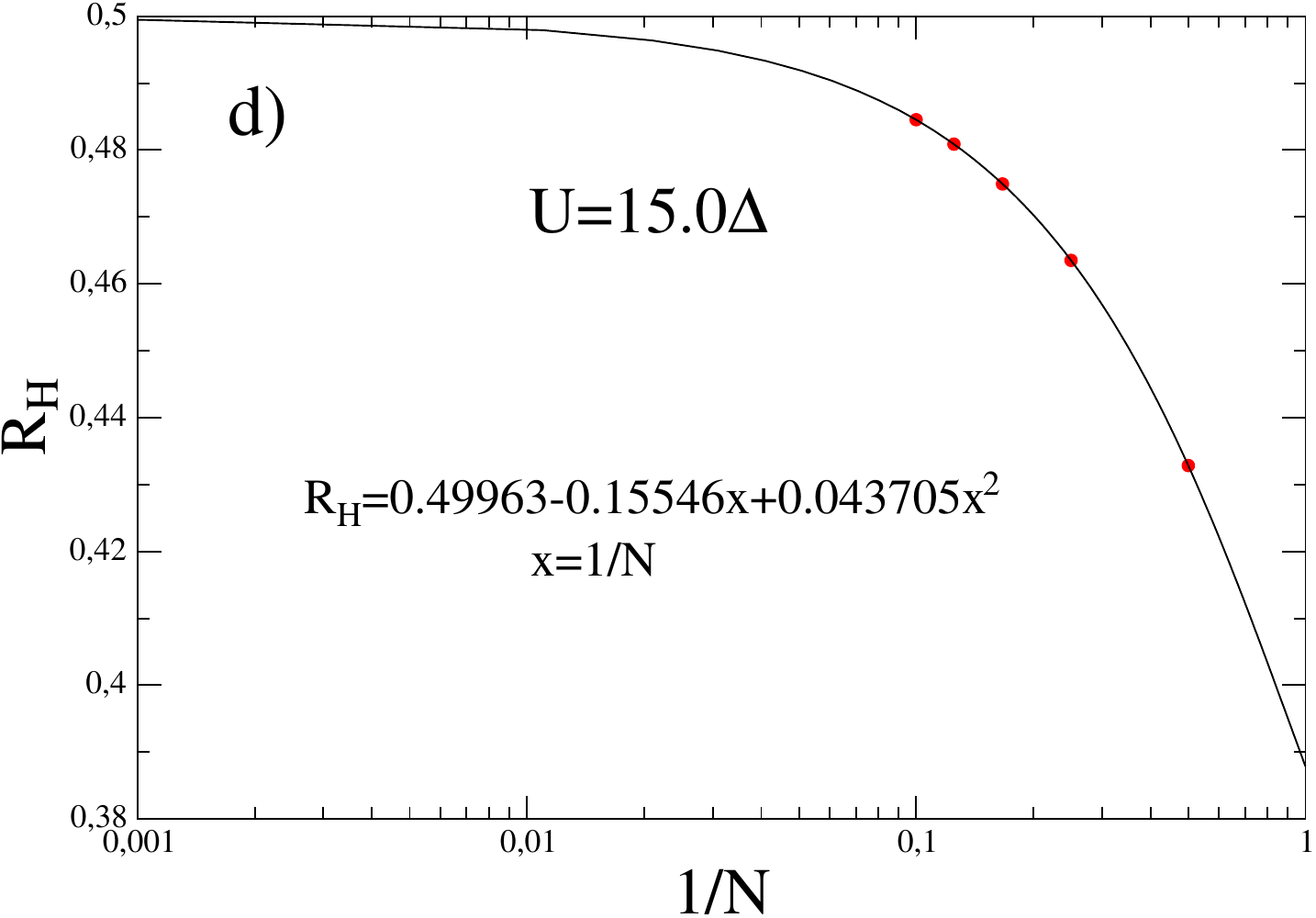}
	\caption{Regression of the energies and residues for  $U=15.0\Delta$.} 
	\label{regress1}
\end{figure}

In our discussion, we will refer the results for $N=8$ in Figs. \ref{transitions} and \ref{Kondo_formation}, which show the essential characteristics of the model's different regimes. 

In Appendix \ref{sec6}, we discussed the second step of the CGFM, which involved calculating the residues and energy differences for all atomic transitions of the N-FALC. This step is computationally intensive but crucial, as it shows that at very low temperatures, only four atomic transitions contribute to generating the entire SIAM density of states, regardless of the number of sites in the N-FALC and the model's parameters and different regimes: Empty orbital, mixed-valence, and Kondo (refer to Fig. \ref{transitions} for details). The remaining transitions can be disregarded due to their minimal residues. However, the contribution of high-energy states must be considered at temperatures above $T_{K}$. Based on this result, the atomic cumulant Green's function can be expressed using a simple formula.
\begin{align}
&\mathbf{g}^{\text{at}}_{\sigma}\left(\omega\right) =g_{1}+g_{13}+g_{31}+g_{3}=\nonumber \\
&\frac{R_{H}}{\omega-E_{H}}+\frac{R_{H}}{\omega+E_{H}}+\frac{R_{K}}{\omega-E_{K}}+\frac{R_{K}}{\omega+E_{K}} ,
\label{atomic_solution}
\end{align}
where $E_{H}$, $E_{K}$, $R_{H}$, and $R_{K}$ are the energies and the residues, numerically calculated for N-FALCs with $N=2,4,6,8,10$. The labels $H$ and $K$ refer to the Hubbard sub-bands and the Kondo peak, respectively.

In Figure \ref{transitions}, we illustrate the transitions for the particle-hole symmetric case with $N=8$, $\mu=0$, and $T = 0.0001 \Delta$. Panel (a) depicts the situation with $U=10.0 \Delta$, panel (b) with $U=20.0 \Delta$, and panel (c) with $U=30.0 \Delta$. Additionally, asymmetric cases are presented with $U=15.0\Delta$ and $E_{f}=-2.5\Delta$ (mixed-valence/Kondo) in panel (d), $E_{f}=0.0\Delta$ (mixed-valence) in panel (e), and $E_{f}=2.5\Delta$ (empty orbital) in panel (f). In the particle-hole symmetric limit, the lower energy transitions on the left and right produce the Kondo peak and the transitions of the first excited states, the Hubbard sub-bands (see, for example, the red ellipsis of Figure \ref{transitions}a). The residues (the numbers next to the transitions) and the transition energies are symmetrical and equal in the symmetric limit but differ for asymmetric cases. Figures \ref{transitions}(a,b,c) indicate that the residues of the Hubbard sub-bands increase with a rise in $U$, while the residues of the Kondo peak decrease, suggesting a reduction in $T_{K}$. Conversely, the difference between the transition energies of the first excited and the ground state widens as $U$ increases, indicating a movement of the Hubbard sub-bands away from the Kondo peak.

\begin{figure}[th]
\centering
\includegraphics[clip,width=0.45\textwidth,angle=0.]{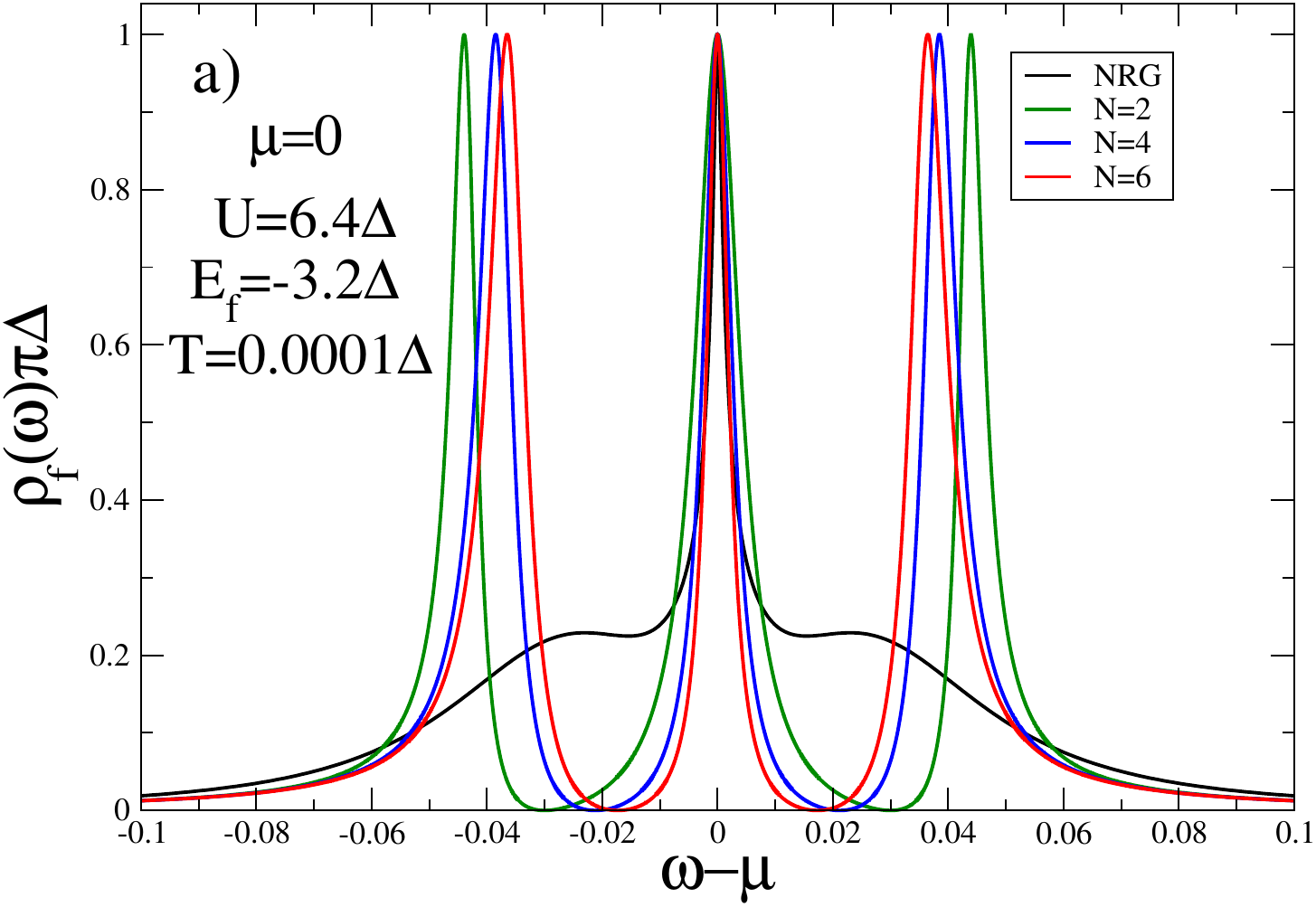}
\includegraphics[clip,width=0.45\textwidth,angle=0.]{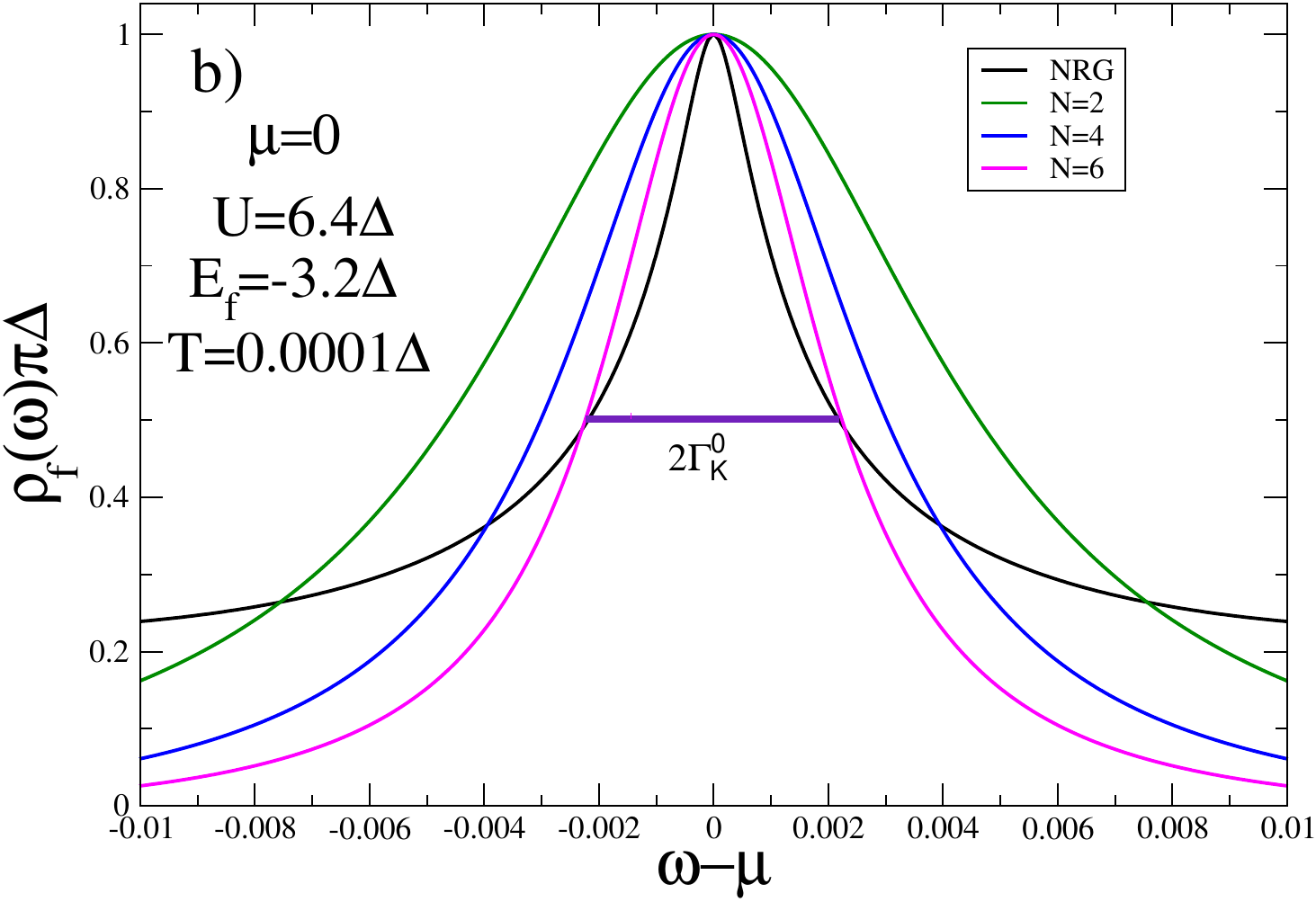}
\includegraphics[clip,width=0.45\textwidth,angle=0.]{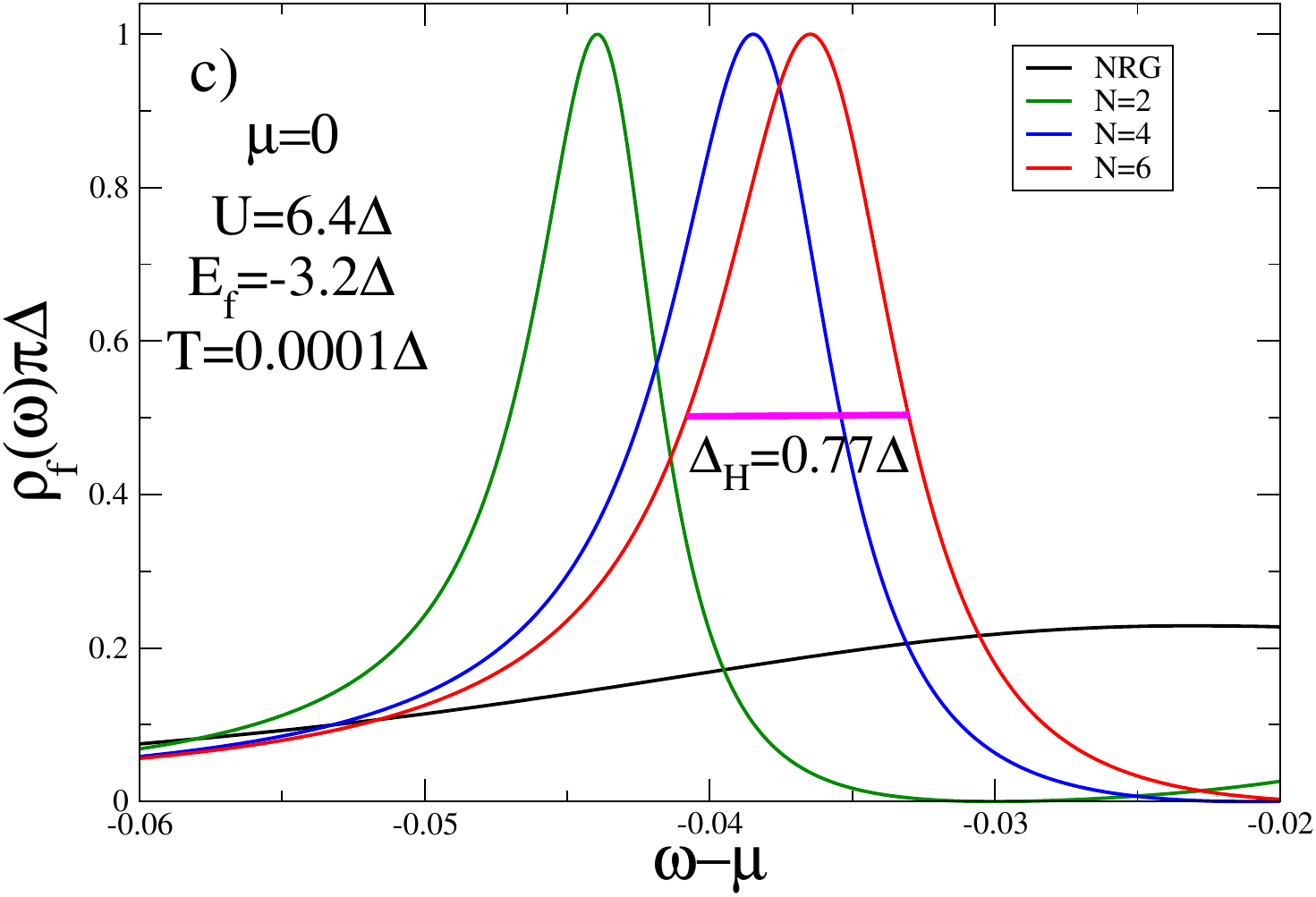}
	\caption{The density of states for the symmetrical limit of the SIAM for $U=6.4\Delta$. a) Full density of states b) Detail of the Kondo peak. c) Detail of the Hubbard sub-bands.}
	\label{Dens64}
\end{figure}

\begin{figure}[th]
\centering
\includegraphics[clip,width=0.45\textwidth,angle=0.]{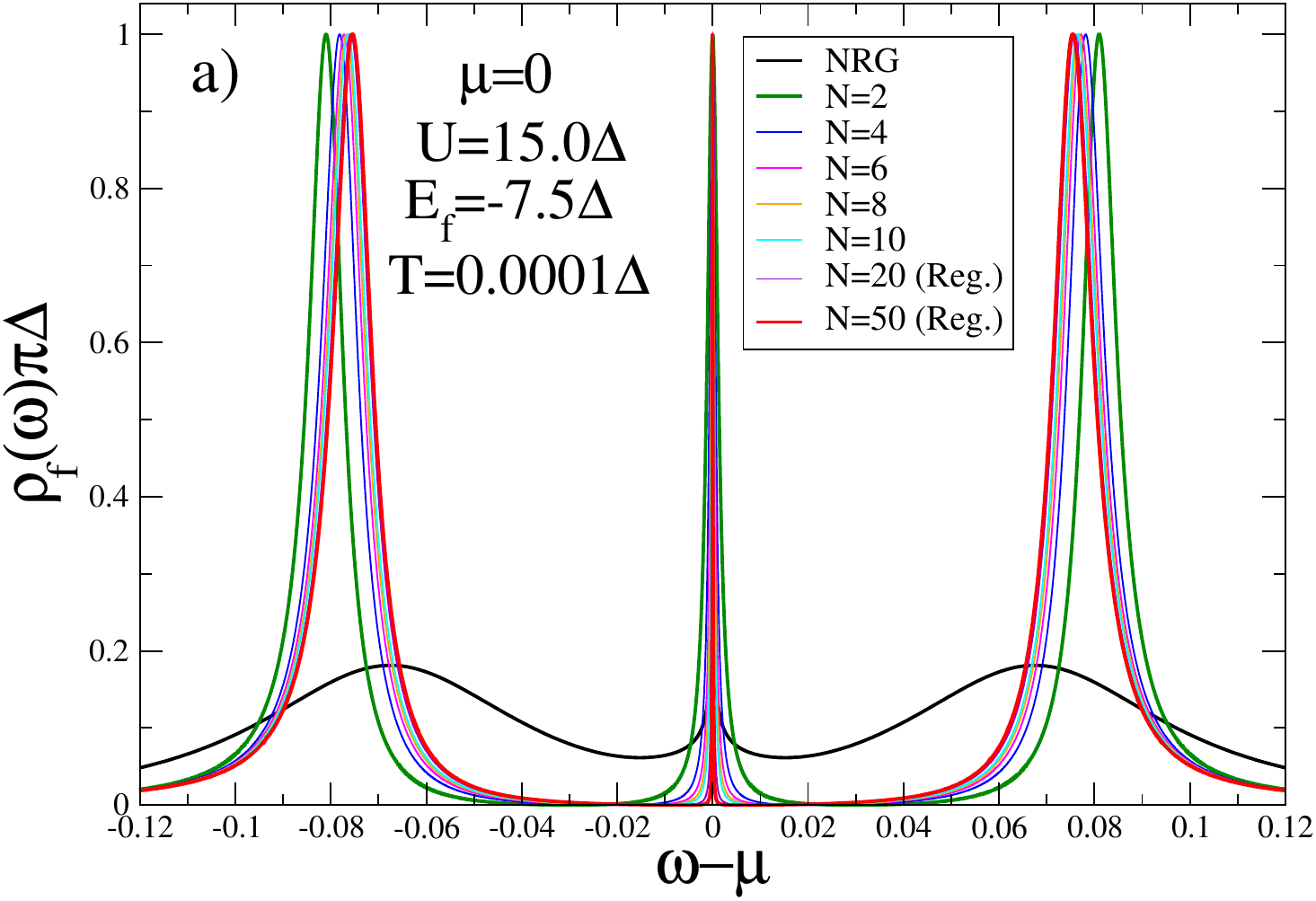}
\includegraphics[clip,width=0.45\textwidth,angle=0.]{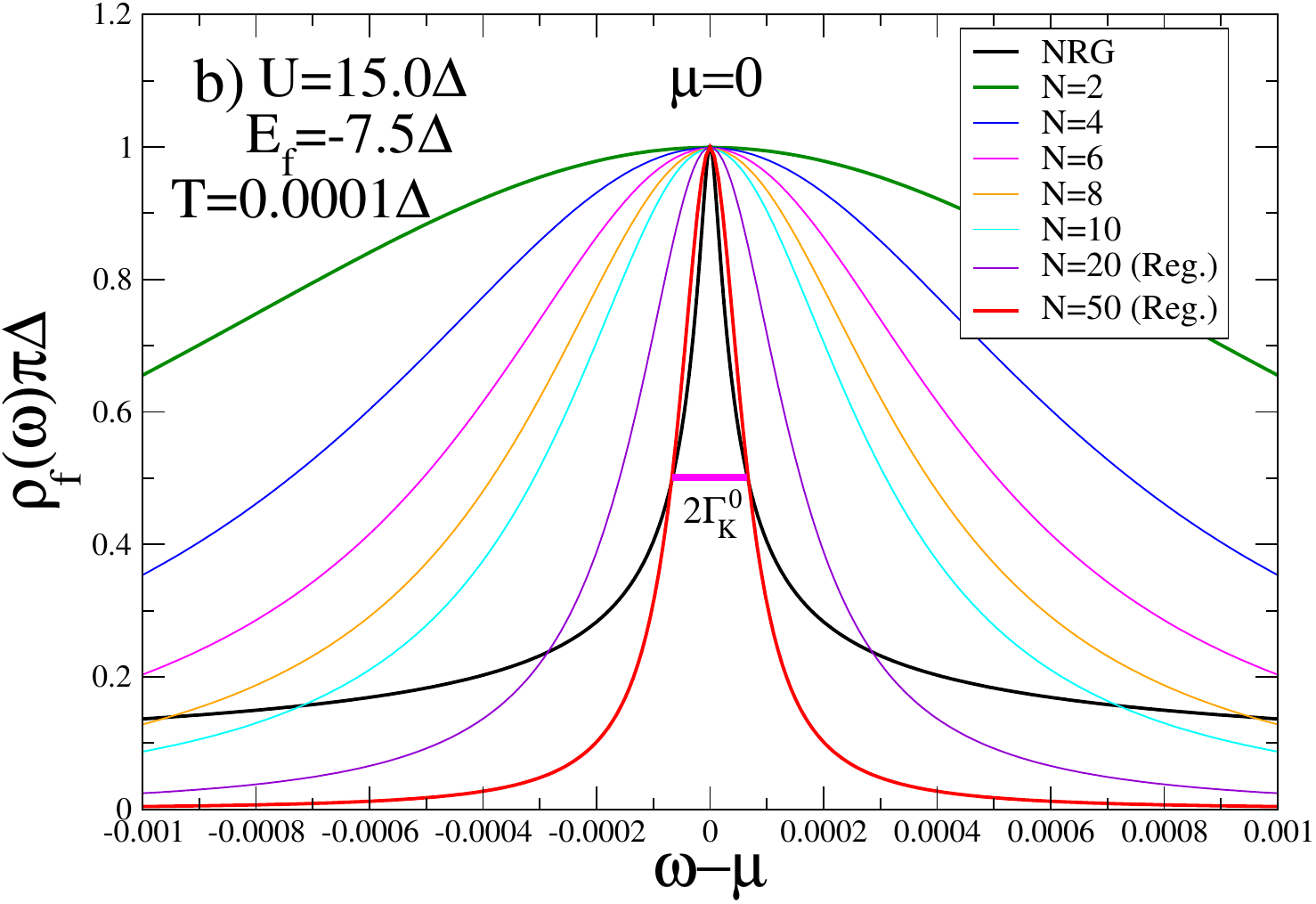}
\includegraphics[clip,width=0.45\textwidth,angle=0.]{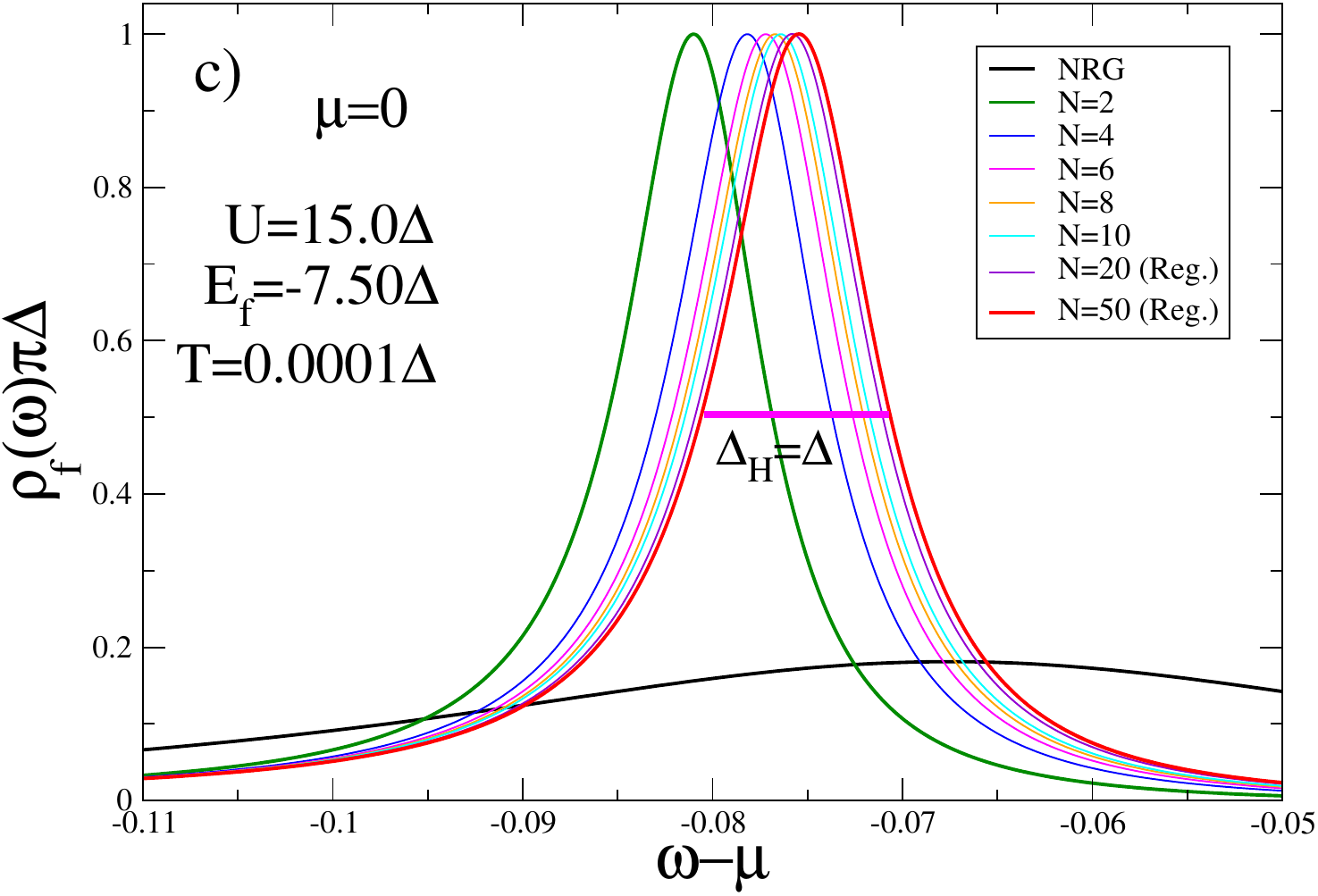}
	\caption{The density of states for the symmetrical limit of the SIAM for $U=15.0\Delta$. a) Full density of states b) Detail of the Kondo peak. c) Detail of the Hubbard sub-bands.}
	\label{Dens15}
\end{figure}

Fig. \ref{Kondo_formation} represents the evolution of the Kondo peak formation corresponding to the energies and residues of the cases d),e) and f) of Fig. \ref{transitions}. In the empty orbital regime (green curve), the density of states is located mainly around the localized energy level at $E_{f}=2.5\Delta$, above $\mu=0$, with a precursor of the upper Hubbard sub-band located at $\omega=0.2$. At $E_{f }=0.0\Delta$, the system enters the mixed-valence regime (blue curve), the localized occupation number increases, the three characteristic structures (the Kondo peak and the Hubbard sub-bands) start to be formed, and the DOS is concentrated around $\mu=0$. In the mixed-valence/Kondo regime (red curve) for $E_{f}=-2.5\Delta$, the Hubbard sub-bands are fully formed, and the Kondo peak starts to be defined. Finally, in the Kondo regime (black curve), at the symmetrical point $E_{f}=-7.5\Delta$, the Hubbard sub-bands are symmetrically about $\mu=0$. The Kondo peak appears at $\mu=0$ (see Figs. (\ref{Dens64} and \ref{Dens15}) (a,b,c) for a discussion of the entire DOS).

The general behavior of the four parameters $E_{H}$, $E_{K}$, $R_{H}$, and $R_{K}$ in Eq. \ref{atomic_solution} for N-FALCs with $N>10$ can be determined using regression analysis. This is illustrated in Fig. \ref{regress1}, which displays the quadratic regression (black curves) obtained from the numerically calculated energies and residues of the Kondo peak and Hubbard sub-bands (red points). These results are based on the exact diagonalization of an N-FALC with $N=2, 4, 6, 8, 10$ for the symmetric limit of the SIAM. We tried different functions to fit the red points and found that the quadratic function fits well up to N=10. However, for $N>10$, we require additional data points to improve these results. The parameters used in the calculation are $U=15.0\Delta$, $\mu=0$, and $T=0.0001\Delta$.

The fact that the density of states, at low temperatures below $T_{K}$, for all regimes of the SIAM, is thoroughly described by only four transitions is a particular characteristic of the SIAM, which probably also happens for other members of the impurity Anderson model's family, like the Kondo and the Coqblin-Schrieffer impurity models. The same does not happen with the Hubbard model, where the number of states that contribute to the density of states is immense, as indicated in one of our previous works of the CGFM (See Fig. 3 of \cite{Lira_JPCM_2023}).


Fig. \ref{Dens64} (a,b,c) shows the CGFM density of states for a particle hole-symmetric limit using the following parameters: $U=6.4\Delta$, $E_{f} = -3.2\Delta$, $\mu=0$, and $T = 0.0001 \Delta$. These parameters were obtained from an experimental realization of a SET system in a mixed-valence/Kondo regime to study its thermoelectric properties \cite{Artis2018}. Other examples of SET experimental results include the original work of Goldhaber-Gordon \cite{Goldhaber1998}, which measured the electrical conductance of a SET at an intermediate Kondo regime with $U=12.9\Delta$, and more recently the work of Bivas Dutta et al. \cite{Dutta2018}, which measured the Seebeck coefficient for a SET in the Kondo limit with $U \simeq 22.0 \Delta$. The CGFM DOS was obtained from the exact diagonalization of the entire Hilbert space generated by a $N=2, 4, 6$-FALC, sufficient to produce a Kondo peak with the same HWHM as the NRG results. The NRG DOS was obtained from the Ljubljana open-source code \cite{Zitko2021}.

In Fig. \ref{Dens64}b), a detailed view of the Kondo peak with the Haldane Kondo temperature $T_{K} \simeq 0.0595\Delta$ is presented. The HWHM of the Kondo peak, which is approximately $\Gamma_{K}^{0} \simeq 3.7 T_{K}=0.222\Delta$, accurately matches the NRG result (black curve) and the CGFM for an N=6-FALC (magenta curve),  is sufficient to recover the NRG HWHM width.

In Fig. \ref{Dens64}(c), a detailed view of the Hubbard sub-band with a HWHM of $\Delta_{H}=0.77\Delta$ is presented. This value does not match the expected result of $\Delta_{H}=\Delta$, indicating that these parameters do not correspond to a well-defined Kondo regime. Instead, they suggest a transition region from a mixed-valence to a Kondo regime. As the number of sites considered in the $N$-FALC increases, the Hubbard sub-bands shift towards the localized energy level $E_{f}$ and energies $E_{f}+U$. On the other hand, the height of the CGFM Hubbard sub-band matches that of the Kondo peak, which is a direct result of the calculations. 

Fig. \ref{Dens15}(a,b,c) shows a fully developed Kondo situation. Fig. \ref{Dens15}a) shows the CGFM density of states for a particle hole-symmetric condition with $T = 0.0001 \Delta$, $E_{d} = -7.5\Delta$, $U=15.0\Delta$, $\mu=0$ and different numbers of sites ($N=2,4,6,8,10$) in the N-FALC. Fig. \ref{Dens15}b) exhibits a detail of the Kondo peak with the Haldane Kondo temperature  $T_{K} \simeq 0.003111\Delta$. The HWHM of the Kondo peak, $\Gamma_{K}^{0} \simeq 3.7 T_{K}=0.0115\Delta$, which is the HWHM of the NRG result (black curve) and the  CGFM results need to advance in the regression process up to $N=50$  to recover the NRG HWHM width. Fig. \ref{Dens15}c) shows that the Hubbard sub-bands are well described by the CGFM, exhibiting the correct width $\Delta_{H}=\Delta$.

\begin{figure}[h]
\centering
\includegraphics[clip,width=0.45\textwidth,angle=0.]{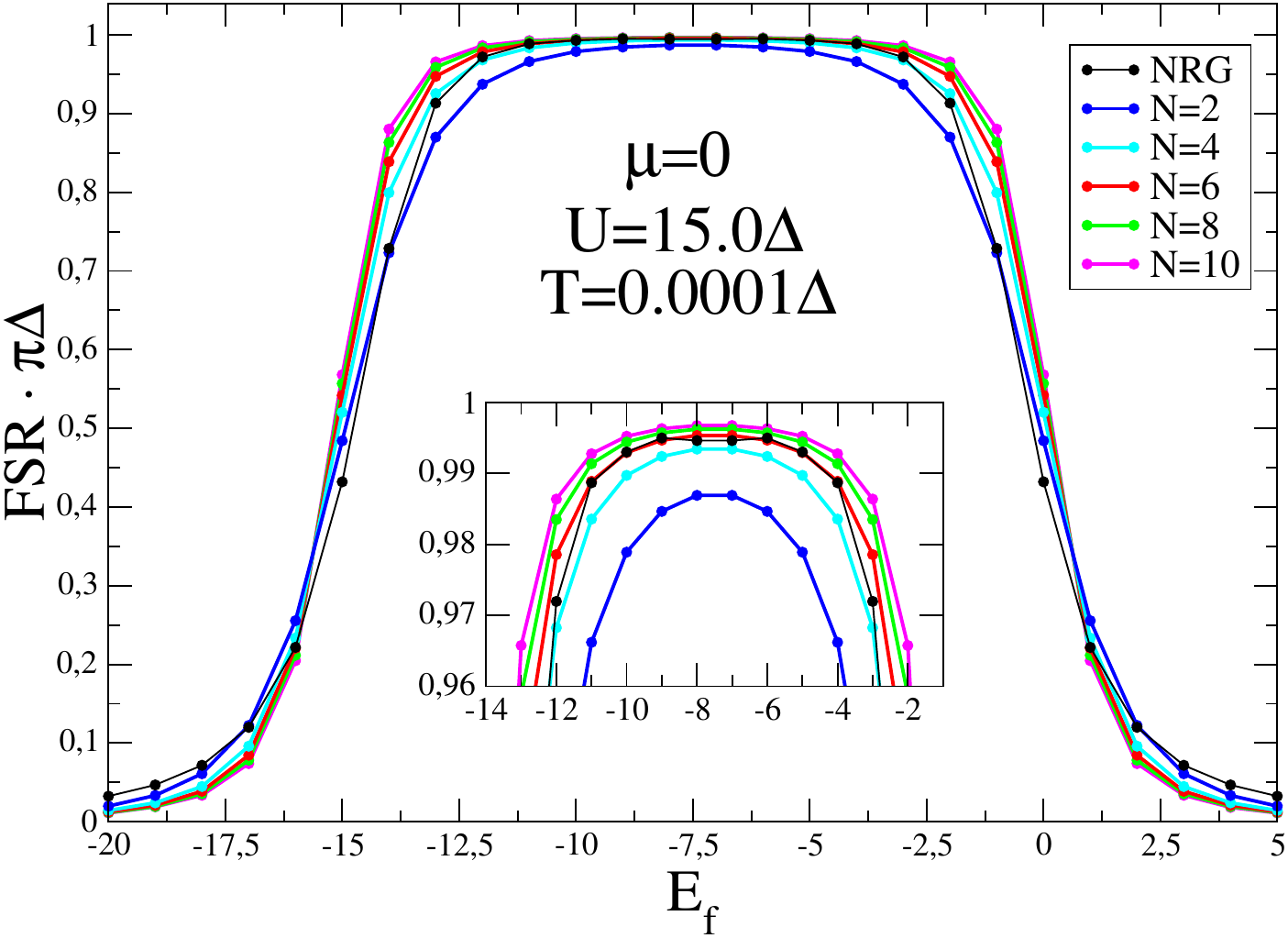}
	\caption{Friedel sum rule as a function of the localized energy level $E_{f}$, for $T = 0.0001 \Delta$, $U=15.0 \Delta$, $\mu=0$ and different numbers of N-FALC sites, ($N=2,4,6,8,10$). The black curve is the NRG result.}
 	\label{fig3}
\end{figure}

Fig. \ref{fig3} shows the FSR as a function of $E_{f}$, for $T = 0.0001 \Delta$, $U=15.0 \Delta$, $\mu=0$ and different numbers of N-FALC sites (N=2,4,6,8,10). The FSR plotted refers to the right side of Eq. \ref{fried} and depends on the localized occupation number $n_{f\sigma}$. The result shows good agreement between the NRG and the CGFM  in the Kondo region. In the CGFM case, all the Hilbert space was solved exactly for the N-FALC until N=10, and the results improved with the growth of N.

\begin{figure}[h]
\centering
\includegraphics[clip,width=0.45\textwidth,angle=0.]{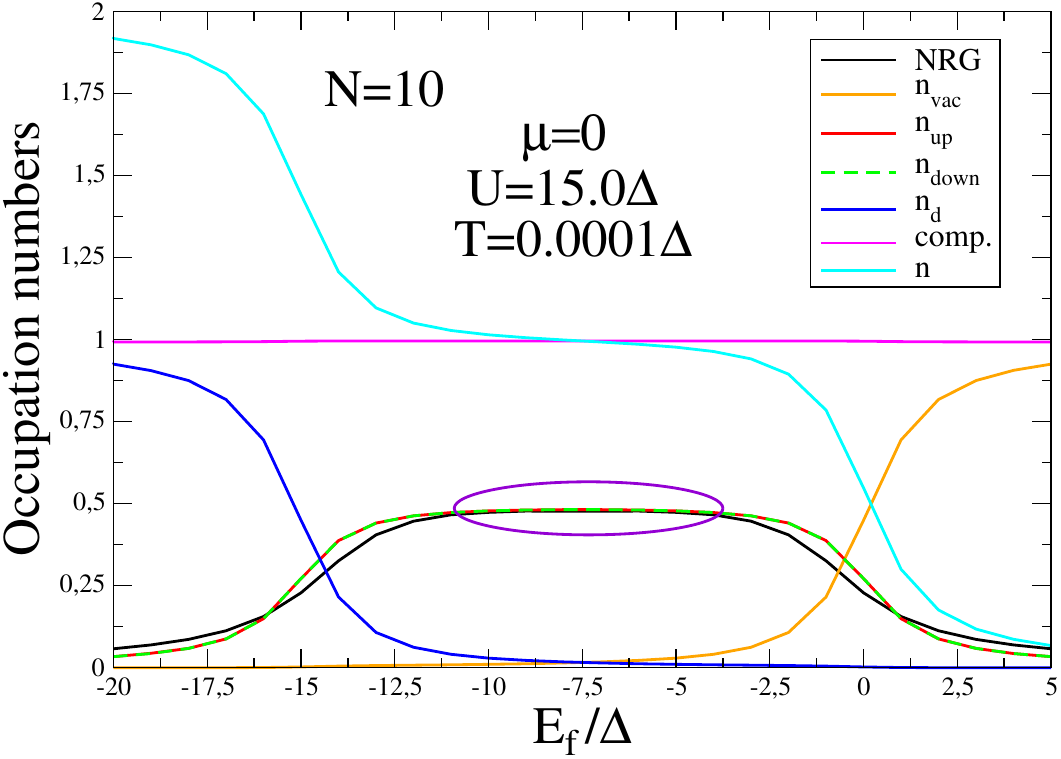}
	\caption{The partial localized occupation number as a function of the localized energy level $E_{f}$,  $U=15.0 \Delta$,  and the $N$-SIALC for $N=10$. The black curve is the NRG result.}
	\label{fig4}
\end{figure}

The CGFM allows for calculating all partial GFs used to determine the occupation numbers. Figure 4 shows the partial localized occupation numbers $n_{vac}$, $n_{up}$, $n_{down}$, and $n_{d}$, as well as the total occupation $n$ and the completeness $comp$ as functions of the localized energy level $E_{f}$, for $T = 0.0001 \Delta$, $U=15.0 \Delta$, $\mu=0$, and the N-FALC for $N=10$. The CGFM adjusts perfectly well with the NRG in the Kondo region, as indicated by the violet ellipsis in the figure.

\section{Conclusions and Perspectives}
\label{sec5}

The CGFM is a method designed to deal with strongly correlated electron systems. In earlier papers \cite{Lira_JPCM_2023, Lira_PLA_2023}, we applied the method to study the one-dimensional single-band Hubbard model and benchmarked the results with the corresponding Bethe ansatz. In this paper, we used the CGFM to investigate the single impurity Anderson model and benchmarked the results with those obtained using the NRG. The CGFM offers another route to addressing strongly correlated electron systems and can be applied to various models such as the Anderson lattice Hamiltonian, the $t-J$, Kondo, Coqblin-Schrieffer, and other related models, across all parameter regions in 1D, 2D, and 3D. The method's calculations are direct, requiring no self-consistency.

We applied the CGFM to the SIAM and calculated the key properties of the model: The density of states in different model regimes with a precise description of the Kondo peak and the calculation of the Friedel sum rule. Another important characteristic of the CGFM is the calculation of the partial GFs associated with the different transition processes as represented in Table \ref{table1}, which allows the calculation of the partial occupation numbers directly, as indicated in Eqs. \ref{G001}-\ref{Gdd1} and Fig. \ref{fig4}. All results were benchmarked with the NRG, and a good agreement was reached. However, one point that remains to be explored more deeply is the quadratic regression used to investigate the Kondo limit, as indicated in Fig. \ref{Dens15}. We need to extend the calculations of the N-FALC to $N=12$ or more to validate the quadratic regression employed in the work.

We have conducted a thorough analysis of the N-FALC up to $N=10$ sites, focusing on the residues of the atomic transitions and their associated energy differences. Our findings reveal that, at very low temperatures, regardless of the parameter settings in different regimes of the model: Empty orbital, mixed-valence, and Kondo, only four atomic transitions are responsible for generating the entire SIAM density of states. Specifically, two of these transitions contribute to the Kondo peak, while the other two generate the Hubbard sub-bands. This unique characteristic of the SIAM contrasts with the Hubbard model, where a multitude of states contribute to the density of states. This is the primary result of our work, consistently observed across all steps of our calculations for $N=2,4,6,8,10$ sites. As a result, we have proposed a four-parameter formula (Eq. \ref{atomic_solution}) to describe the atomic cumulant GFs. Furthermore, we have utilized regression analysis to extend our findings to cases with $N>10$, enabling us to benchmark our density of states results with the NRG.

The result suggests that one potential approach to finding the exact solution of the SIAM is to study the energy level structure of the N-FALC Hilbert space in order to identify the states that contribute to producing the four final atomic transitions. It is essential to carry out the diagonalization process from the beginning, starting with N=2 and continuing as N increases, in order to identify which states will contribute to the ground state and the first excited state that gives rise to the Kondo peak and the Hubbard sub-bands. We plan to explore this further in the future.

\begin{acknowledgments}
M.S.F. acknowledges financial support from the Brazilian National Council for Scientific and Technological Development (CNPq) Grant. Nr. 311980/2021-0 and to Foundation for Support of Research in the State of Rio de Janeiro (FAPERJ) process Nr. 210 355/2018. 
\end{acknowledgments}

\appendix

\section{The cumulant Green's function method}
\label{sec6}

In this Appendix, we will briefly discuss the fundamental relations of the CGFM. For more details,  refer to the following references: \cite{Figueira1994, Foglio2010, Lobo2010, Lira_JPCM_2023, Lira_PLA_2023}. As mentioned in the Introduction, we will outline the basic concepts of the CGFM in four steps:

(i) We will begin by calculating the eigenenergies and eigenvectors of the N-FALC using ED methods.

(ii) Next, we will use the Lehmann representation to calculate all the atomic Green's functions associated with possible transitions within the N-FALC Hilbert space.

(iii) Utilizing these atomic Green's functions, we will calculate the atomic cumulants.

(iv) Finally, we will use these cumulants to approximate the full SIAM GFs.

Using the Eq. \ref{E2.8a}, the GF for the localized electrons with spin $\sigma$ can be calculated through the $4 \times 4$ matrix equation \cite{Foglio2010}
\begin{equation}
\mathbf{G}_{\sigma }^{f}(z)=\mathbf{M}_{\sigma}(z)\mathbf{\cdot }\left( \mathbf{%
I-A}_{\sigma}(z)\right) ^{-1},  \label{exactGF}
\end{equation}%
and, from this equation it follows 
\begin{equation}
\mathbf{M}_{\sigma }(z)\mathbf{=}\left[ \mathbf{I+G}_{\sigma }^{f}(z)\cdot 
\mathbf{W}_{\sigma }(z)\right]^{-1}\cdot \mathbf{G}_{\sigma }^{f}(z),
\label{exactM}
\end{equation}
where $\mathbf{M}_\sigma$ is the cumulant and $\mathbf{A}_\sigma=\mathbf{W}\cdot \mathbf{M}$. If the Hamiltonian is spin independent, or commutes with the z component of the spin, the above matrices can be reduced to two $2\times2$ matrices, where $\mathbf{W}$ is defined as 
\begin{eqnarray}
\mathbf{W}_{\uparrow }\left( z\right) &=&\left\vert V\right\vert ^{2}\varphi
_{\uparrow }(z)%
\begin{pmatrix}
1 & 1 \\ 
1 & 1%Dens64
\end{pmatrix}%
,  \label{E5.5a} \\
\mathbf{W}_{\downarrow }\left( z\right) &=&\left\vert V\right\vert
^{2}\varphi _{\downarrow }(z)\ 
\begin{pmatrix}
1 & -1 \\ 
-1 & 1%
\end{pmatrix}
.
\end{eqnarray}
The function $\varphi_{\sigma}(z)$ represents the conduction band and it is equal to
\begin{equation}
\varphi _{\sigma }(z)=\frac{1}{N_{s}}\sum_{\mathbf{k}}\ G_{\sigma }^{0c}\left( \mathbf{k},z\right),  \label{E5.6a}
\end{equation}%
where 
$G_{\sigma }^{0c}\left(\mathbf{k},z\right)=[z-\varepsilon \left( \mathbf{k},\sigma \right)]^{-1}$ 
is the GF for a free particle representing the conduction electrons.

Considering the uncorrelated rectangular conduction band of semi-width $D=1$,
\begin{equation} 
\label{SB1}
\rho_{c}^{0}( \epsilon_{\bf k}) =
\left\{
    \begin{array}{l}
     \frac{1}{2D} \; 
     ~~~~\mbox{ for } -D \le \epsilon_{\bf k} \le D\\
      ~0 ~~\;  ~~~~\mbox{ otherwise }
    \end{array}  \right. .
\end{equation}

The Anderson parameter is given by 
\begin{equation}
\Delta=\pi V^{2} \rho_{c}^{0}({\mu=0})=\frac{1}{2D} . 
\label{Delta}
\end{equation}

Integrating  Eq.~\eqref{E5.6a}  we obtain
\begin{equation}
\varphi _{\sigma }(z)=\frac{1}{2D}\ \ln \left( \frac{z-D+\mu }{z+D+\mu }\right) .  
\label{E5.6}
\end{equation}

In the first step, we use ED methods to numerically solve the N-FALC equation (Eq. \ref{Linearchain}) for $N=2,4,6,8,10$. Initially, we solved it for the first two sites, which form a dimer (as described in Lobo et al., \cite{Lobo2010}). One site represents localized electrons, and the other represents conduction electrons. We then include a second conduction site in the chain by performing the tensor product between its states and the dimer states. This process is repeated for a third conduction site, and so on, until the desired N value is reached. The Hilbert space dimension for an N-FALC scales as $4^{N}$, as shown in Table~\ref{table1}. Additionally, we utilize the symmetries of the SIAM to transform the Hamiltonian matrix into its block diagonal form. The process of diagonalizing N-FALCs for $N=2,4,6,8$ is carried out using the Fortran 90 BLAS/LAPACK routines.

Next, in the second step, using the eigenvalues and the eigenvectors obtained in the previous step, we calculate the atomic GFs, employing the Lehmann spectral representation~\cite{Fetter1971}
\begin{equation}
\mathcal{G}_{\alpha\alpha^{\prime}}^{ff,at}(i\omega_{s})=-e^{\beta\Omega}%
\sum_{n,r,r^{\prime}}\frac{\exp(-\beta\varepsilon_{n-1,r})+\exp(-\beta
\varepsilon_{n,r^{\prime}})}{i\omega_{s}+\varepsilon_{n-1,r}-\varepsilon
_{n,r^{\prime}}}\times  \notag
\end{equation}
\begin{equation}
\times \left\langle n-1,r\right\vert\ X_{j,\alpha}\left\vert
n,r^{\prime}\right\rangle \left\langle n,r^{\prime}\right\vert \ X_{j,\alpha
^{\prime}}^{\dagger}\ \left\vert n-1,r\right\rangle ,  \label{EqLehmann}
\end{equation}
where $\Omega=-k_{B}T ln {\cal{Z}}$ is the grand-canonical potential ($K_{B}$ is the 
Boltzmann constant) and 
${\cal{Z}}=\sum_{n,r} \exp{-\beta \varepsilon_{n,r}}$ is the grand-canonical ensemble 
partition function. Finally, $\varepsilon_{n,r} = E_{n,r}-n\mu$, where $E_{n,r}$ is 
the energy of the $r$ state $|n, r>$, with $n$ indicating the number of electrons present 
in the state and $\mu$ is the chemical potential. 

The most challenging part of the computation for the CGFM is the numerical calculation of the allowed transitions within the Hilbert space of N-FALC. This task consumes significant time due to the necessity of calculating the allowed transitions' spectral weights (residues). Transitions with zero residues or residues less than a specified threshold (in our case, $10^{-9}$) are disregarded, and the remaining transitions are utilized to compute the atomic Green's functions (GFs).

For $N > 8$,  we use the main result of the work: At low temperatures, only four atomic transitions are responsible for generating the entire SIAM density of states. To compute the eigenenergies and eigenvectors associated with these four states, we utilize the FEAST Library package \cite{Polizzi2009, Tang2014}. While this approach has proven effective, it is essential to first generate the entire Hilbert space with $4^{10}$ states before utilizing the FEAST package to search for states within the low-energy region. The results obtained employing the CGFM  depend upon the available computational resources. The Fortran code for $N=10$ was running on a workstation with 256 GB of RAM and a high-performance SSD. However, more powerful hardware is necessary for  N-FALCs with $N>10$. After calculating the transitions and residues, the atomic GF can be expressed as:
\begin{equation}  \label{GAnderson}
\mathbf{g}^{at}_{\sigma}(z) = e^{\beta\Omega} \sum_{i=1}^{M} \frac{m_{i}}{z-u_{i}%
},
\end{equation}
where the $u_{i}$ and $m_i$ are the poles and residues of the atomic GF, respectively, and $M$ is the number of transitions kept, i.e., transitions with residues greater than the threshold established for the calculations.

We should calculate the atomic GFs considering that only the elementary processes associated with the electron spin annihilation (creation) are allowed for possible atomic transitions, as indicated in the tables \ref{table1} and \ref{tab:3}. The GFs $g_{1}$, $g_{13}$, $g_{31}$ and $g_{3}$ are associated with the transitions that annihilate (create) a spin-up electron while $g_{2}$, $g_{24}$, $g_{42}$ and $g_{4}$ are related to the creation (annihilation) of a spin-down electron. According to Table~\ref{tab:3}, it can be seen that the GFs $g_{1}$ and $g_{2}$ represent states that initially contain only one electron, whereas the GFs $g_{3}$ and $g_{4}$ are associated to states with two electrons. On the other hand, $g_{13}$, $g_{31}$, $g_{24}$ and $g_{42}$ are the crossing GFs, which contain simultaneous annihilation (creation) of electrons in states with singly and doubly occupied states. Thus, the atomic Green's function can be written as 
\begin{equation}
\label{eq:771}
\mathbf{g}^{at}_{\sigma}(z) = \left( \begin{array}{cccc}
g_{1} & g_{13}  & 0 & 0 \\
g_{31}  & g_{3}  & 0 & 0 \\
0 & 0 & g_{2}  & g_{24}  \\
0 & 0 & g_{42}  & g_{4} 
\end{array} \right).
\end{equation}
Equation \eqref{eq:771} is block-diagonal because the selection rules do not allow spin-flip transitions, i.e., spin up and down are disconnected.

In the third step, to calculate the atomic cumulants, we substitute the atomic GFs obtained in the second step, Eq.~\eqref{eq:771}, into the corresponding Dyson equation, Eq. \eqref{exactM}, and we arrive at
\begin{equation}
\mathbf{m}^{at}_{\sigma }(z)\mathbf{=}\left[ \mathbf{I+g}_{\sigma }^{at}(z)\cdot 
\mathbf{W}^{o}_{\sigma }(z)\right]^{-1}\cdot \mathbf{g}_{\sigma }^{at}(z),
\label{approxM}
\end{equation}
where the $\mathbf{W}_{\sigma }^{o}$ is equal to 
\begin{align}
\mathbf{W}_{\uparrow }^{o}\left( z\right) & =\left\vert \Delta \right\vert
^{2}\varphi _{\uparrow }^{o}(z)%
\begin{pmatrix}
1 & 1 \\ 
1 & 1%
\end{pmatrix}%
,  \label{E5.55a} \\
\mathbf{W}_{\downarrow }^{o}\left( z\right) & =\left\vert \Delta \right\vert
^{2}\varphi _{\downarrow }^{o}(z)\ 
\begin{pmatrix}
1 & -1 \\ 
-1 & 1%
\end{pmatrix}
,  \label{E5.55b}
\end{align}
and
\begin{equation}
\varphi _{\sigma }^{o}(z)=\frac{-1}{z-E_{q}-\mu },
\label{Eq3.144}
\end{equation}%
is the GF of the uncorrelated conduction atomic sites, with the N-FALC energy $E_{q}=\epsilon_{n}=0$. In Eqs. \ref{E5.55a} and \ref{E5.55b}, we substitute the real hybridization $V$ of the entire conduction band by a reduced hybridization defined by the Anderson parameter $\Delta$. When using the N-FALC, we reduce the whole conduction Hamiltonian to a small set of conduction atomic sites. This process leads to a concentration of all the conduction electrons at the atomic level $\epsilon_{n}=0$ of the conduction sites, which overestimates their contribution to the effective cumulant. We found numerically that the reduced hybridization should be the Anderson parameter, $\Delta$, which produces a well-defined Kondo peak at the chemical potential.
\begin{table}
\caption{Atomic Green's functions for the finite $U$ SIAM Hamiltonian: $\alpha=(b,a)$ represents a transition from an initial state $a$ to a final state $b$.}
\centering
{\setlength{\extrarowheight}{3.0pt}
{\renewcommand{\arraystretch}{1.0}
\begin{tabular}{ c  c  c  c  c }
\toprule
$g^{\text{at}}_{\uparrow}$& $g_{1}$ & $g_{3}$ & $g_{13}$ & $g_{31}$ \\
\midrule
$(b,a)$ & $(0,\uparrow)$ & $(\downarrow,d)$ & $(0,\uparrow)$ and $(\downarrow,d)$ & $(\downarrow,d)$ and $(0,\uparrow)$ \\
\midrule
$g^{\text{at}}_{\downarrow}$ & $g_{2}$ & $g_{4}$ & $g_{24}$ & $g_{42}$ \\
\midrule
$(b,a)$ & $(0,\downarrow)$ & $(\uparrow,d)$ & $(0,\downarrow)$ and $(\uparrow,d)$ & $(\uparrow,d)$ and $(0,\downarrow)$ \\
\bottomrule
\end{tabular}}}
\label{tab:3}
\end{table}

The  cumulants $\mathbf{m}^{at}_{\sigma }(z)$ are calculated, using the Eqs. \ref{eq:771}-\ref{Eq3.144}, as an aproximation of the exact cumulants
\begin{equation}
\mathbf{M}_{\uparrow}(z)=
\begin{pmatrix}
M_{1} & M_{13} \\
M_{31} & M_{3}%
\end{pmatrix}= \notag
\begin{pmatrix}
m^{at}_{1} & m^{at}_{13} \\
m^{at}_{31} & m^{at}_{3}%
\end{pmatrix}= \notag
\end{equation}
\begin{equation}
=\frac{%
 \begin{pmatrix}
g_{1} & g_{13} \\
g_{31} & g_{3}
\end{pmatrix} 
+\left\vert \Delta\right\vert ^{2}\varphi_{\uparrow}^{o}(z)\left(
g_{1} g_{3} - g_{13} g_{31}\right)
\begin{pmatrix}
1 & -1 \\
-1 & 1%
\end{pmatrix}
}{1+\left\vert \Delta \right\vert ^{2}\varphi_{\uparrow}^{o}(z)\left(
g_{1} + g_{3} + g_{13} + g_{31}\right) } , \label{5.121}
\end{equation}
\begin{equation}
\mathbf{M}_{\downarrow}(z)=\begin{pmatrix}
M_{2} & M_{24} \\
M_{42} & M_{4}%
\end{pmatrix} = \notag
\begin{pmatrix}
m^{at}_{ 2} & m^{at}_{24} \\
m^{at}_{42} & m^{at}_{4}%
\end{pmatrix}= \notag
\end{equation}
\begin{equation}
\frac{%
\begin{pmatrix}
g_{2} & g_{24} \\
g_{42} & g_{4}
\end{pmatrix}
+\left\vert \Delta \right\vert ^{2}\varphi_{\downarrow}^{o}(z)\left(
g_{2} g_{4} - g_{24} g_{42}\right)
\begin{pmatrix}
1 & 1 \\
1 & 1%
\end{pmatrix}
}{1+\left\vert \Delta \right\vert ^{2}\varphi_{\downarrow}^{o}(z)\left(
g_{2} + g_{4} - g_{24}- g_{42}\right) } . \label{5.122}
\end{equation}

In the fourth step of the CGFM, we calculate all the full SIAM GFs. Substituting the approximate cumulants obtained in Eqs. \ref{5.121}-\ref{5.122} into Eq. \ref{exactGF} and carrying out the analytical continuation from the pure imaginary Matsubara GFs, $z=i\omega=\omega+i\eta$, with $\eta=10^{-6}$ in our calculations, to the real frequency axis, we obtain the entire localized GF
\begin{equation}
\mathbf{G}_{\uparrow}^{f}(\omega)=\begin{pmatrix}
G^{f}_{1} & G^{f}_{13} \\
G^{f}_{31} & G^{f}_{3}%
\end{pmatrix} = \notag \\
\end{equation}
\begin{equation}
\frac{%
\begin{pmatrix}
M_{1} & M_{13} \\
M_{31} & M _{3}%
\end{pmatrix}
-\left\vert V\right\vert ^{2}\varphi_{\uparrow}(\omega)
\Theta_{13}
\begin{pmatrix}
1 & -1 \\
-1 & 1%
\end{pmatrix}
}{1-\left\vert V\right\vert ^{2}\varphi_{\uparrow}(\omega)
\Gamma_{13}} ,  \label{E5.12}
\end{equation}
\begin{equation}
\mathbf{G}_{\downarrow}^{f}(\omega)=\begin{pmatrix}
G^{f}_{2} & G^{f}_{24} \\
G^{f}_{42} & G^{f}_{4}%
\end{pmatrix} = \notag \\
\end{equation}
\begin{equation}
\frac{%
\begin{pmatrix}
M_{2} & M_{24} \\
M_{42} & M_{4}%
\end{pmatrix}
-\left\vert V\right\vert ^{2}\varphi_{\downarrow}(\omega)\Theta_{24}
\begin{pmatrix}
1 & 1 \\
1 & 1%
\end{pmatrix}
}{1-\left\vert V\right\vert ^{2}\varphi_{\downarrow}(\omega)\Gamma_{24}} ,  \label{E5.13}
\end{equation}
where $\varphi_{\sigma}(\omega)$ is given by Eq. \ref{E5.6}. $\Theta_{13}=M_{1}M_{3}-M_{13}M_{31}$, $\Gamma_{13}=M_{1}+M_{13}+M_{31}+M_{3}$, $\Theta_{24}=M_{2}M_{4}-M_{24}M_{42}$ and $\Gamma_{24}=M_{2}+M_{24}-M_{42}-M_{4}$.

In the same way, from Eq. \ref{E2.8b} we can obtain the conduction GF, $\mathbf{G}_{\sigma}^{c}(\omega)$. A detailed derivation can be found in \cite{Foglio2010}. The electronic correlations appear in the Eqs. through the cumulants GFs: $\Gamma_{13}$ and $\Gamma_{24}$. The  conduction GFs for spins up and down are given by
\begin{equation}
\mathbf{G}_{\uparrow}^{c}(\omega)=\frac{\varphi_{\sigma}(z)}{1-\left\vert
V\right\vert ^{2}\varphi_{\sigma}(\omega)\Gamma_{13}} ,
\label{E5.36}
\end{equation}
\begin{equation}
\mathbf{G}_{\downarrow}^{c}(\omega)=\frac{\varphi_{\sigma}(z)}{1-\left\vert
V\right\vert ^{2}\varphi_{\sigma}(\omega) \Gamma_{24}} .
\label{E5.36}
\end{equation}

The cross Green function $\mathbf{G}_{\sigma}^{cf}(\omega)$ is defined by a column vector with two components as \cite{Foglio2010}
\begin{equation}
G_{\sigma}^{cf}(\omega)=
\begin{pmatrix}
G_{\uparrow}^{cf}(\omega)\\
G_{\downarrow}^{cf}(\omega)
\end{pmatrix} ,
\end{equation}

\begin{equation}
\mathbf{G}_{\uparrow}^{cf}(\omega)=-V\frac{\varphi_{\uparrow}(i\omega
)}{1-\left\vert V\right\vert ^{2}\varphi_{\uparrow}(i\omega)\Gamma_{13}}
\begin{pmatrix}
m_{11}+m_{31}\\
m_{13}+m_{33} 
\end{pmatrix} ,
\label{E5.34}%
\end{equation}
and%
\begin{equation}
\mathbf{G}_{\downarrow}^{cf}(\omega)=-V\frac{\varphi_{\downarrow}
(i\omega)}{1-\left\vert V\right\vert ^{2}\varphi_{\downarrow}(i\omega
)\Gamma_{24}}
\begin{pmatrix}
m_{22}-m_{42}\\
m_{24}-m_{44}
\end{pmatrix} .
\label{E5.35}%
\end{equation}

\bibliography{References_Anderson.bib}

\end{document}